\documentclass[twocolumn]{aastex631}
\usepackage{amsmath}

\newcommand{\ie}{\emph{i.e.}}
\newcommand{\eg}{\emph{e.g.}}

\newcommand{\dM}{\ensuremath{\mathrm{d}M}}
\newcommand{\angstrom}{\text{\normalfont\AA}}
\newcommand{\dDCR}{\ensuremath{\Delta\mathrm{DCR}}}
\usepackage{hyperref}

\begin{document}

\title{Every Datapoint Counts: Stellar Flares as a Case Study of Atmosphere Aided Studies of Transients in the LSST Era}

\author[0000-0001-9273-5036]{Riley W. Clarke}
\affiliation{Department of Physics \& Astronomy,
University of Delaware,
Newark, DE 19716, USA}
\affiliation{Data Science Institute,
University of Delaware,
Newark, DE 19716, USA}

\author[0000-0002-0637-835X]{James R. A. Davenport}
\affiliation{DiRAC Institute, Department of Astronomy,
University of Washington,
Seattle, WA 98195, USA}

\author[0000-0002-8916-1972]{John Gizis}
\affiliation{Department of Physics \& Astronomy,
University of Delaware,
Newark, DE 19716, USA}
\affiliation{Data Science Institute,
University of Delaware,
Newark, DE 19716, USA}

\author[0000-0002-9154-3136]{Melissa L. Graham}
\affiliation{DiRAC Institute, Department of Astronomy,
University of Washington,
Seattle, WA 98195, USA}

\author[0000-0002-0514-5650]{Xiaolong Li}
\affiliation{Department of Physics \& Astronomy,
University of Delaware,
Newark, DE 19716, USA}
\affiliation{Data Science Institute,
University of Delaware,
Newark, DE 19716, USA}

\author[0000-0001-7559-7890]{Willow Fortino}
\affiliation{Department of Physics \& Astronomy,
University of Delaware,
Newark, DE 19716, USA}
\affiliation{Data Science Institute,
University of Delaware,
Newark, DE 19716, USA}

\author{Ian Sullivan}
\affiliation{DiRAC Institute, Department of Astronomy,
University of Washington,
Seattle, WA 98195, USA}

\author{Yusra Alsayyad}
\affiliation{Department of Astrophysical Sciences, Princeton University, Princeton, NJ, 08544, USA}

\author[0000-0003-2759-5764]{James Bosch}
\affiliation{Department of Astrophysical Sciences, Princeton University, Princeton, NJ, 08544, USA}

\author{Robert A. Knop}
\affiliation{E.O. Lawrence Berkeley National Laboratory, 1 Cyclotron Rd., Berkeley, CA, 94720, USA}

\author[0000-0003-1953-8727]{Federica Bianco}
\affiliation{Department of Physics \& Astronomy,
University of Delaware,
Newark, DE 19716, USA}
\affiliation{Data Science Institute,
University of Delaware,
Newark, DE 19716, USA}
\affiliation{Joseph R. Biden School of Public Policy and Administration,
University of Delaware,
Newark, DE 19716, USA}
\affiliation{Vera C. Rubin Observatory}



\begin{abstract}

Due to their short timescale, stellar flares are a challenging target for the most modern synoptic sky surveys. The upcoming Vera C. Rubin Legacy Survey of Space and Time (LSST), a project designed to collect more data than any precursor survey, is unlikely to detect flares with more than one data point in its main survey. We developed a methodology to enable LSST studies of stellar flares, with a focus on flare temperature and temperature evolution, which remain poorly constrained compared to flare morphology. By leveraging the sensitivity expected from the Rubin system, Differential Chromatic Refraction can be used to constrain flare temperature from a single-epoch detection, which will enable statistical studies of flare temperatures and constrain models of the physical processes behind flare emission using the unprecedentedly high volume of data produced by Rubin over the 10-year LSST. We model the refraction effect as a function of the atmospheric column density, photometric filter, and temperature of the flare, and show that flare temperatures at or above $\sim$4,000$K$ can be constrained by a single $g$-band observation at airmass $X\gtrsim1.2$, given the minimum specified requirement on single-visit relative astrometric accuracy of LSST, and that a surprisingly large number of LSST observations is in fact likely be conducted at $X\gtrsim1.2$, in spite of image quality requirements pushing the survey to preferentially low $X$. Having failed to measure flare DCR in LSST precursor surveys, we make recommendations on survey design and data products that enable these studies in LSST and other future surveys.
\end{abstract}



\section{Introduction} \label{sec:intro}

Differential Chromatic Refraction (DCR) is the apparent shift of a celestial object on the sky due to atmospheric refraction. The magnitude of the shift is determined by the airmass, observing bandpass, and Spectral Energy Distribution (SED, which is dominated by the temperature and spectral composition) of the source. The direction of the shift is toward the zenith. 
DCR must be characterized effectively to ensure it does not compromise image quality, astrometric inference \citep{1967pras.book.....V}, spectrophotometry \citep{filippenko1982}, and in surveys where time-domain phenomena are discovered and studied in template subtracted images \citep{lupton2007}. DCR, however, can also aid scientific inference in a few cases. For example, \citet{kaczmarczik2009} used DCR to improve redshift measurements of quasars in the Sloan Digital Sky Survey (SDSS). While innovative techniques have been previously employed to study the chromatic evolution of stellar flares (\eg\ \citet{hedges2021}), the use of ground-based observations to infer information about flare SEDs via atmospheric refraction remains untested. We consider flares to be suitable candidates for inferential applications of DCR due to their stochastic and short-lived nature, which makes detections in multiple filters difficult, their nature as chromatic transients, and their heightened occurrence on cool stars like M dwarfs, the most common type of star in our galaxy. Currently, there are only a small number of measurements of flare temperatures, which our usage of DCR will enable.

Flares are stochastic, short-lived stellar transients caused by magnetic reconnection at the stellar photosphere \citep{pettersen1989}, most commonly on low-mass stars such as M-dwarfs (\dM\ hereafter). Flares are tracers of stellar magnetic activity, and flare rates have been studied as a correlate with stellar age and rotation \citep{davenport2016}. Flares can also have a significant impact on planetary atmospheres, as high amounts of UV flux from flares can deplete their ozone layers \citep{tilley2019} and are thought to have a role in the formation of life, where they have been identified as both a potential trigger and inhibitor \citep{ramsay2021}. 

Photometrically, they appear as a highly chromatic sudden rise (typically $\sim 3$ magnitudes in $u$ band and between 0.5 and 1 magnitude in $g$ band, see \citealt{davenport2012}) followed by an exponential decay phase on a timescale ranging from a few minutes to 100-200 minutes in some cases \citep{yan2021}. While the photometric evolution of flares is well-understood thanks to high-accuracy, rapid-cadence measurements by exoplanet-finding satellites like \textit{Kepler} \citep{borucki2003} and \textit{TESS} \citep{ricker2010}, accurate color measurements remain scarce. Statistical inference on flares' origin and evolution mechanisms and their impact is limited by this lack of a large ensemble of information on flare temperatures and temperature evolution, 
prompting a turn to large next-generation astrophysical surveys, most notably the Vera C. Rubin Observatory Legacy Survey of Space (LSST).

LSST \citep{ivezic2019} is the premier ground-based photometric survey of the 2020s. It will continuously scan the whole of the southern sky over a 10-year period, producing over 3 million images with a sub-arcsecond resolution of the whole southern sky in six photometric filters: $ugrizy$. Based on flare rate measurements in SDSS Stripe 82 \citep{kowalski2009}, the number of flares will range between 0.4 and 1.4 per image, depending on galactic latitude.
While it is certain that Rubin will observe many flares, whether or not we can access scientifically valuable information from those detections is currently an open question. It is expected that the LSST primary survey, referred to as ``Wide, Fast, Deep'' (WFD), will take 2-3 exposures per pointing per night with a median internight gap of 34 minutes, repeating observations of a field in the same filter every few days ($\sim6$ days in $r$, $\sim21$ days in $g$, median values, see \citealt{PSTN-055} for details on the most recent LSST observing plans). Given that {\citealt{yan2021} reported median rise and decay times of flares on Sun-like stars in the \textit{Kepler} catalog to be 5.9 minutes to 22.6 minutes respectively, LSST will observe most flares in the WFD with 1-2 data points per detected event, rendering the time-resolved photometric analysis enjoyed by \eg\ Kepler impossible. The Deep Drilling Fields (DDF) will offer an opportunity for time-resolved flares and to identify even faint \dM\ flare precursors. Roughly 5\% of the total survey time will be devoted to continuous observations of five selected pointings. While the observational details for the DDFs are yet to be defined, in one hour over 100 consecutive exposures can be taken, over 20 in each of five filters of one 9.6 sq degree field of view.\footnote{Note that the Rubin filter wheel can host 5 filters on any night. The reddest and bluest filters are swapped based on moon cycles and five filters are available when observing.}  When co-added with the main survey observations these images will extend the 5$\sigma$ depth down to $r\sim 27.5$, allowing time-resolved observations of even the faintest \dM s in multiple filters. It must be noted that all DDFs are extragalactic fields, where the flare rate is expected to be suppressed.\footnote{For details on the selection of the fields see \citealt{PSTN-055}.} Yet flares are common phenomena even in the extragalactic sky, and we expect hundreds to be detected in each DDF by the end of the 10-year survey}. The LSST DDF program will enable traditional flare studies. 
However, by exploiting the high astrometric accuracy of Rubin, here we show that the Rubin WFD, with its unprecedented sample size, offers a valuable opportunity for flare science.

It is already expected that DCR will aid extragalactic studies of quasars \citep{yu2020, richards2018} with Rubin, and we here propose that employing DCR as a scientific tool can be extended to variable and transient phenomena with Rubin. We use stellar flares as our case study and demonstrate in this paper that by taking advantage of Rubin’s pristine image quality and astrometric precision to measure the difference in source location on the sky between quiescence and event due to DCR, we will be able to indirectly obtain information regarding the color, and therefore the temperature of the flare, from even a single point detection. We will show that an astrometric shift should be apparent in LSST astrometry for a star when flaring compared to when quiescent, with the effect being especially pronounced for a hot flare on a cool star observed at high airmass; but we further demonstrate that the effect will be measurable by Rubin even at its typically low airmass range. This means that, despite the sparse photometric sampling of Rubin, the DCR toolset can be used to leverage its high volume of flare detections in order to infer flare temperatures across an expansive sample. 

In \autoref{sec:method} we describe how DCR is calculated and how we model the expected excess DCR during a flare with some temperature, at some airmass, in a given filter. In \autoref{sec:rubincapability} we compare the expected DCR for flaring stars with the capabilities of the Rubin system and data analysis pipeline. In  \autoref{sec:precursor} we describe our failed search for excess DCR produced by flares in surveys considered precursors to LSST, ZTF \citep{bellm2018} and Deep Drilling in the Time Domain with DECam \citep{graham2023} and identify the bottlenecks that impaired a successful DCR measurement. In  \autoref{sec:recommend} we discuss the technical requirements for measuring the \dDCR\ from flares: the excess zenith-bound displacement of a flaring star compared to the apparent shift expected based on its quiescent SED, and provide recommendations to maximize the potential of this method for LSST and other surveys. In  \autoref{sec:conclusions} we summarize our conclusions and explain how this method can be embedded within Rubin's software ecosystem to support studies of flares and other chromatic phenomena throughout the survey lifespan.

\section{Methodology} \label{sec:method}
\begin{figure}[t!]
    \centering
    \includegraphics[width=0.48\textwidth]{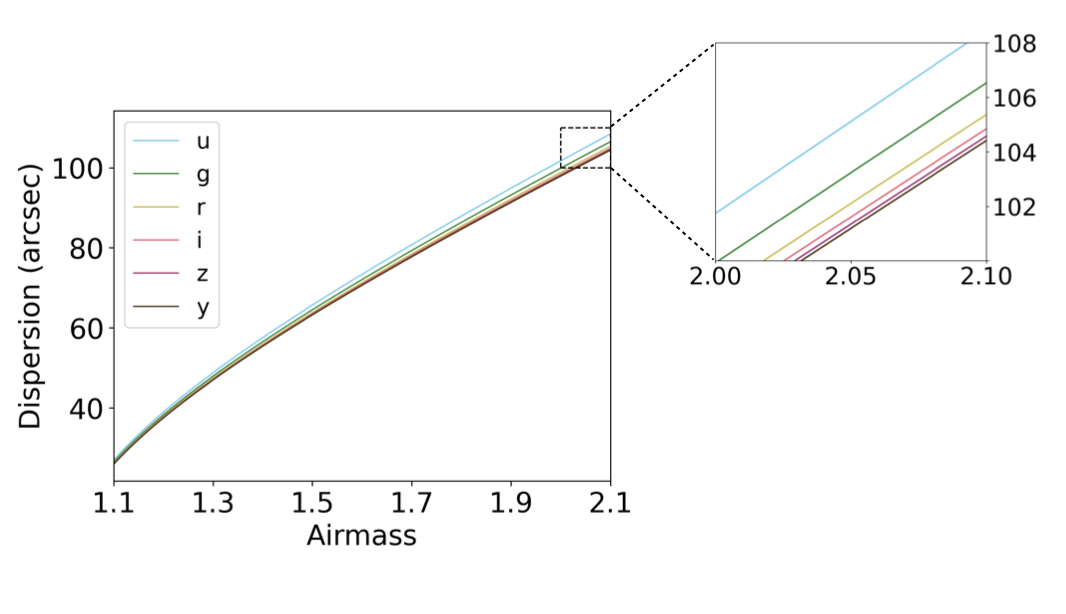}
    \caption{Angular deflection from true position for a source with a 10,000$K$ blackbody spectrum for airmass $1.1\leq X\leq2.1$ in all six LSST bands. A zoomed section is shown to reveal the separation between curves corresponding to separate bandpasses.}
    \label{fig:dcr_diagram2}
\end{figure}
  
  Following \citep{kaczmarczik2009}, we can describe the formalism of DCR in the following four steps:

\begin{equation}
    \lambda_\mathrm{eff} = \frac{\int_0^\infty f_\lambda S_j(\lambda)\ln(\lambda)d\lambda}{\int_0^\infty f_\lambda S_j(\lambda)d\lambda}
    \label{eq:1}
\end{equation}

\noindent
defines $\lambda_\mathrm{eff}$ as the effective wavelength of bandpass $j$, where $S_j$ is the transmission function of $j$, and $f_\lambda$ is the spectral flux of the source (or SED) in units of $Wm^{-2}\mu^{-1}$; for an observed source then, the filter-dependent refraction index $n_\lambda$ can be calculated from
\begin{equation}
\begin{split}
[n_\lambda - 1]~10^6 = 64.328\: + & \frac{29498.1}{146-(1/\lambda_\mathrm{eff})^2} \\
                              + & \frac{255.4}{41 - (1/\lambda_\mathrm{eff})^2};
\end{split}
\end{equation}

\noindent
$R$, the total angular deflection of the source due to DCR (in arcseconds), is then calculated as:

\begin{equation}
    R_0(\lambda) = \frac{n_\lambda^2 - 1}{2n_\lambda^2},
\end{equation}

\begin{equation}
    R = R_0(\lambda)\tan(Z)
\end{equation}

\noindent
where Z is the zenith angle, and where the airmass $X$ is commonly approximated as $X = \sec(Z)$, which assumes a homogenous, plane-parallel atmosphere. This approximation is valid only for $60^{\circ}\le Z\le75^{\circ}$, so our calculations involving airmass will be restricted to $X\le2.1$ (although we note that the LSST survey strategy includes observations at higher airmasses, see \autoref{sec:rubincapability} and \autoref{sec:recommendopsim}). \autoref{fig:dcr_diagram2} illustrates the amplitude of DCR produced by a 10,000$K$ blackbody source for a range of airmasses in the six LSST bandpasses.

\begin{figure}[ht!]
    \centering
    \includegraphics[width=0.5\textwidth]{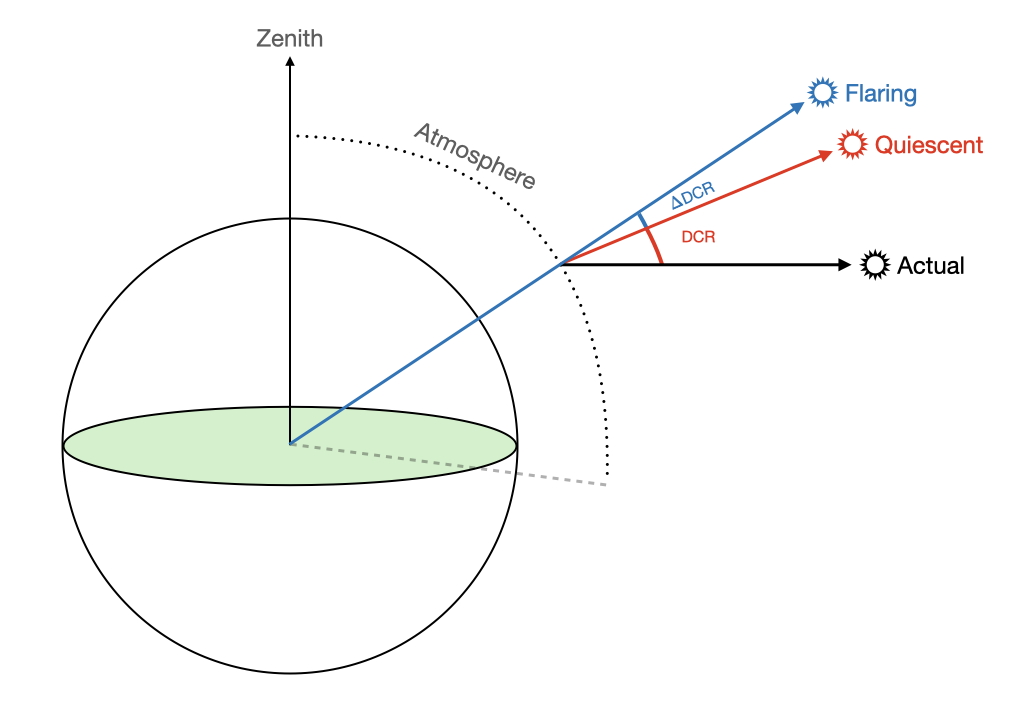}
    \caption{Light incident from a star is deflected by the atmosphere. The amount of deflection depends on the color (\ie\ temperature) of the source and the amount of atmosphere the light passes through. The chromatic change during a flare event should produce an excess in the normal DCR at quiescence, labeled as \dDCR\ in the figure.}
    \label{fig:dcr_diagram1}
\end{figure}

\begin{figure}[bh!]
    \centering
    \includegraphics[width=0.48\textwidth]{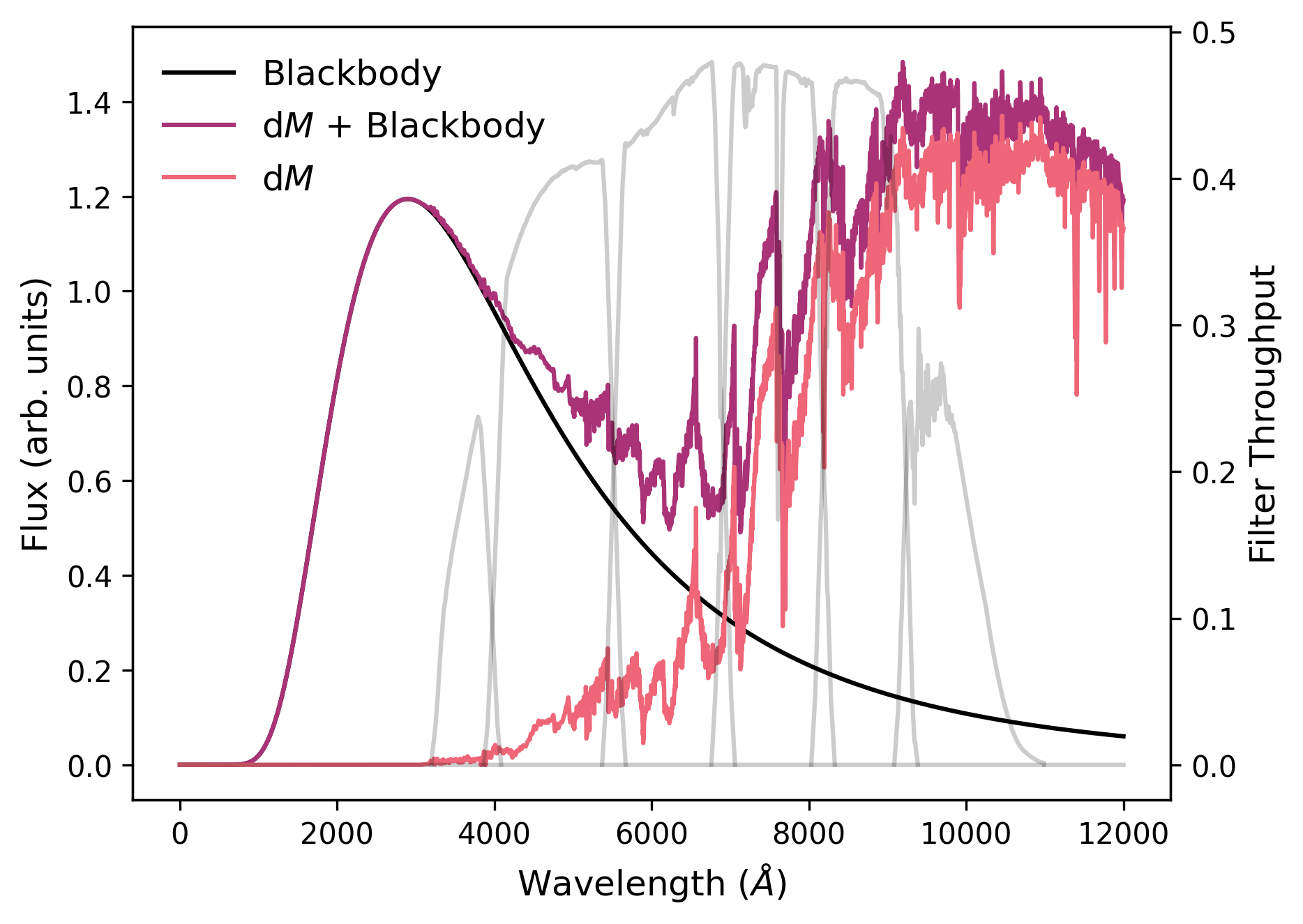}
    \caption{In pink, the spectrum of an M5 dwarf (composite spectrum built partially from SDSS observations, \citealt{davenport2012}). 
    In blue, the spectrum of a 10,000$K$ blackbody to simulate the flare. In purple, the sum of the two aforementioned spectra, representing the spectrum of the \dM\ during the flare event. The blackbody and \dM\ spectra contain the same total energy over the SDSS optical range ($3,850-9,200~\angstrom$). The transmission functions for the LSST $ugrizy$ photometric system are shown in grey.}
    \label{fig:bbcalib}
\end{figure}

In \autoref{sec:precursor} we will demonstrate that even flares with moderately high temperatures observed at airmasses $1.05\leq X\leq1.2$ will produce an astrometric shift that will be detectable at the LSST precision level.

\subsection{The DCR effect induced by flares: \dDCR}

Using the DCR formalism described in \autoref{sec:method}, we created an expository model to illustrate the DCR produced by a flare. We assume the star is an M5 dwarf and model the quiescent SED of the source with a template spectrum spanning the optical and near-infrared wavelength ranges built by \citet{davenport2012}, with the optical component using observations from the Sloan Digital Sky Survey (SDSS). We clip this template spectrum at $12,000\angstrom$ as this was sufficient for full coverage of the LSST $ugrizy$ bands. The source in its quiescent state will already suffer from a DCR effect, but we expect that the temperature contrast between flare and quiescent states will determine a significant change in the apparent star position. The difference between the displacement in quiescent and event state is the \textit{excess} displacement produced by the flare which we refer to as \dDCR. A schematic illustration of \dDCR\ is shown in \autoref{fig:dcr_diagram1}. 

To simulate a flare, we add a blackbody spectrum at temperature T to the spectrum of the star (\autoref{fig:bbcalib}). In the absence of a well-characterized spectral energy distribution (SED), flare spectra are canonically approximated as a 9,000-10,000$K$ blackbody \citep{osten2015}. Thus, in this simulation  the spectra are normalized such that the blackbody component at the canonical flare temperature ($T=10,000$K) has the same total energy as the \dM\ spectrum when both are integrated over the SDSS optical range ($3,850-9,200~\angstrom$). We also explored the effect of changing the fraction of the stellar surface covered by the flaring region, referred to as a ``filling factor''. Assuming our energy-based calibration to be representative of a filling factor  $f_f\sim0.05$ and applying a flat scaling to the blackbody spectrum for other  $f_f$ values, we tested filling factors ranging from 0.05 to 0.2 and found that we are not sensitive to this parameters for $f_f\geq0.05$. The composite spectrum is convolved with the chosen LSST filter to calculate $\lambda_\mathrm{eff}$ in \autoref{eq:1}, allowing us to estimate the angular displacement from DCR during both quiescence and event, and their difference, or $\dDCR$. 

\subsection{Measuring \dDCR\ with the Parallactic Angle}

Many astrophysical and observational effects may cause the apparent position of a star to change between two images of the same area of the sky. Although stars' proper motion is typically too slow to have a measurable effect in images collected within days, weeks, or months, images are however subject to various distortion effects such that a star may appear to be offset between two images, especially if observed in a different position of the CCD plane, with different telescope rotation, \emph{etc}. These observational systematics should be corrected, and we will return to these spurious effects in \autoref{sec:precursor}. These components of the star's displacement should have no preferential direction while the DCR-induced displacement should be strictly in the direction of the zenith. To determine the DCR-induced change in position of a given source between detections, we will measure the components of the apparent motion toward, tangential to, or away from the zenith direction. 
The parallactic angle is defined as the angle between two great circles, one passing through the source and the zenith, and the other passing through the source and the North Celestial Pole. It is calculated according to \citet{meeus1998} as: 

\begin{equation}
    P = \tan^{-1}\left(\frac{\sin(h)}{\cos(\delta)\tan(\phi) - \sin(\delta)\cos(h)}\right),
\end{equation}

\noindent
where $\delta$ is the object's declination, $h$ is the object's hour angle, and $\phi$ is the geographic latitude of the observer's location on Earth. The component of the source's motion in the direction of the zenith (which we call $d_{\parallel}$) is then calculated as:

\begin{equation}
    d_{\parallel} = \sqrt{\Delta\alpha^2 + \Delta\delta^2}\cos\left(\frac{\pi}{2} - P_2 - \tan^{-1}\left(\frac{\Delta\delta}{\Delta\alpha}\right)\right),
\end{equation}

\noindent
where $\alpha$ is the object's right ascension, and $P_2$ is the parallactic angle of the source in its second position. 

\section{Detection Potential for LSST} \label{sec:rubincapability}

\begin{figure}[t!]
    \centering
    \includegraphics[width=0.48\textwidth]{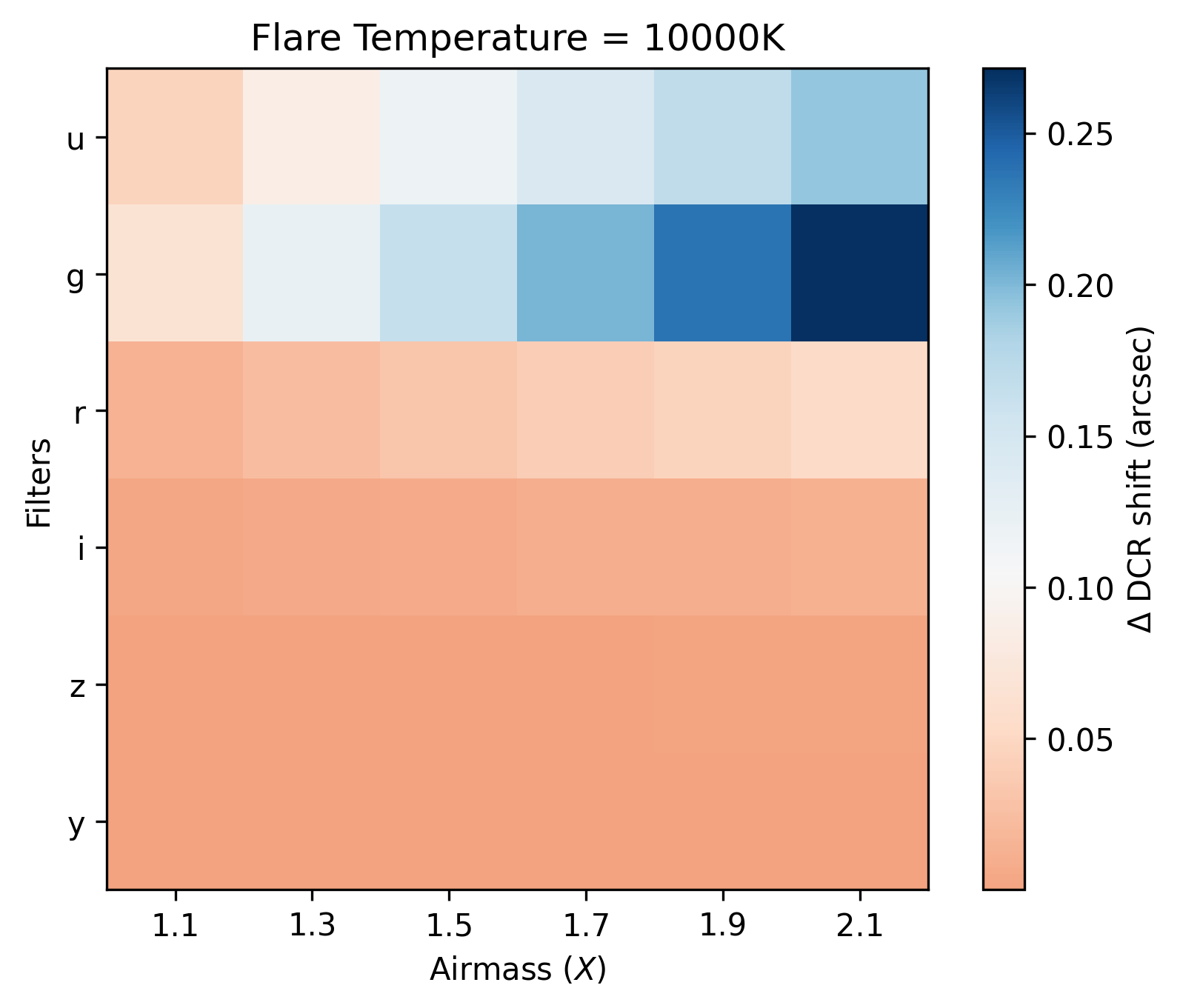}
     \caption{Expected magnitude of the \dDCR\ effect for a flare SED approximated by a 10,000$K$ blackbody as a function of airmass and filter in the Rubin Observatory $ugrizy$ observing system \citep{10.1117/12.790264}. Blue coloring corresponds to a \dDCR\ shift detectable by Rubin, and red coloring corresponds to an undetectable shift, given the absolute astrometric accuracy goal of 0.1 arcsec.}
    \label{fig:deltagrid}
\end{figure}

Using the methodology described in \autoref{sec:method}, we simulate the \dDCR\ produced by a flare on an M5 star for a variety of airmasses and flare temperatures across all six Rubin bands. The \dDCR\ for a fixed flare temperature at 10,000$K$ for all 6 filters in the 1.1-2.1 airmass range is shown in \autoref{fig:deltagrid}. The Rubin system specification requires a single-image absolute astrometric accuracy of Rubin of 0.1 arcsec \citep{LPM-17}. This figure demonstrates that a 10,000$K$ flare's \dDCR\ is detectable in $g$-bands even at moderate airmasses; 
$\sim8\%$ of all WFD images will be in $g$ band. 

While the effect is also detectable in $u$-band, higher airmasses are required to generate a comparably large shift. This may be counterintuitive, as the DCR effect is more pronounced in bluer wavelengths. However, the \dDCR\ is proportional to the change in $\lambda_\mathrm{eff}$ during the flare compared to quiescent state. The quiescent SED has a prominent redward slope in the $g$ band wavelengths which, integrated through the filter leads to a $\lambda_{\mathrm{eff,quiescent}}(g)=4,988.4\angstrom$ which becomes $\lambda_{\mathrm{eff, flare}}(g)=4,740.9\angstrom$ when the black body contribution dominates the SED during flare. Conversely, the quiescent SED in $u$ band is more flat and only slopes redward significantly where the filter transmission is already very low. Thus the shift in $\lambda_\mathrm{eff}$ is smaller ($\lambda_{\mathrm{eff,quiescent}}(u)=3,728.0\angstrom$ compared to $\lambda_{\mathrm{eff,flare}}(u)=3,655.8\angstrom$).

\begin{figure}[ht!]
    \centering
    \includegraphics[width=0.48\textwidth]{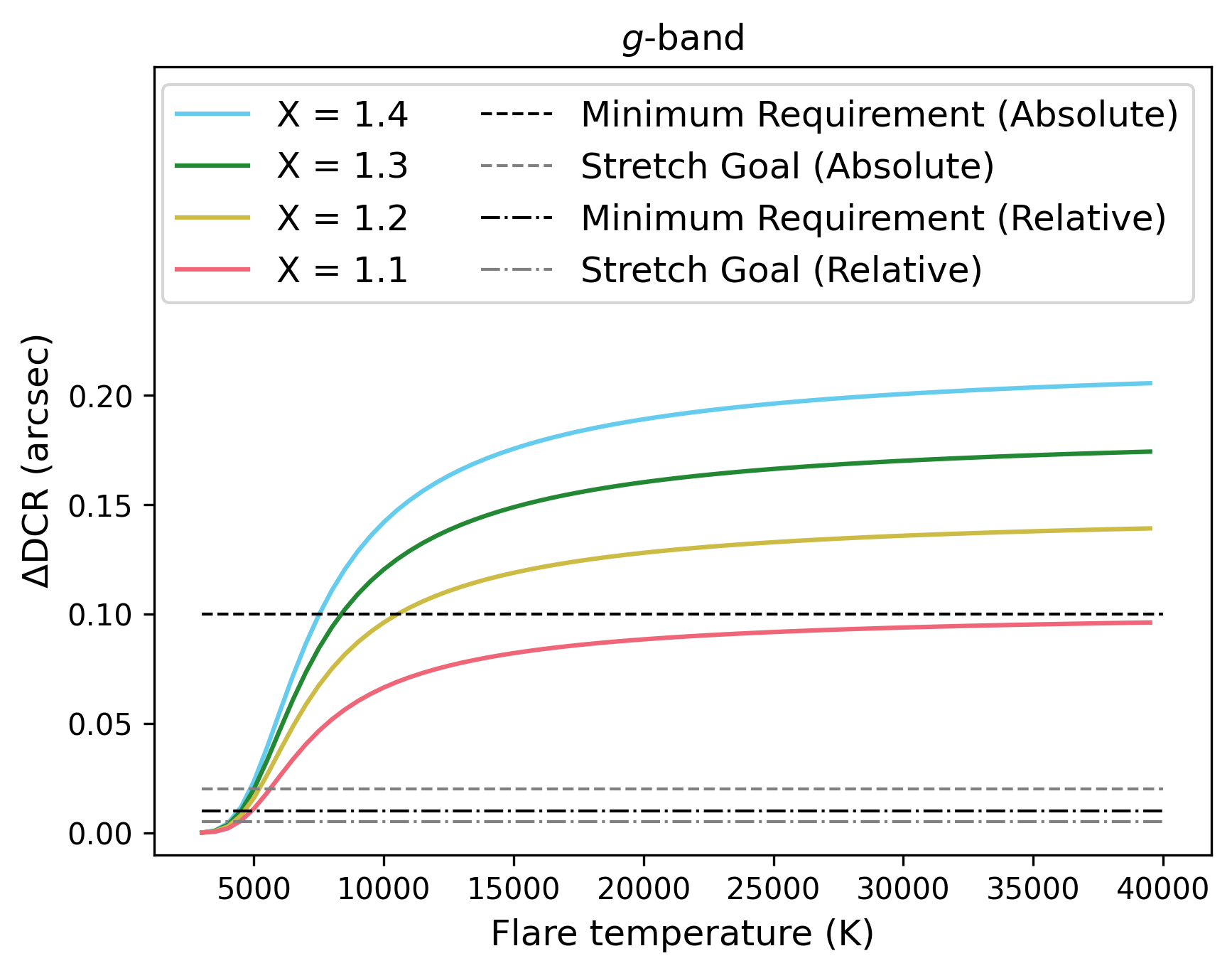}
    \caption{The magnitude of a star image displacement during a flare compared to the quiescent star position as a function of flare temperature in LSST $g$-band for four different airmass values. The minimum requirement and stretch goal for absolute astrometric accuracy of Rubin are shown by the the black and grey dashed lines, respectively. The minimum requirement and stretch goal for relative astrometric accuracy of Rubin are shown by the black and grey dash-dot lines, respectively.}
    \label{fig:deltashifttemp}
\end{figure}

The $g$-band evolution of the \dDCR\ with flare temperature is shown in \autoref{fig:deltashifttemp}, showing that the \dDCR\ grows rapidly between 5,000-10,000$K$, then flattens off at higher temperatures. The minimum system requirement and stretch goals for both absolute and relatve astrometric accuracy of Rubin is indicated on \autoref{fig:deltashifttemp}, demonstrating that DCR could be used to probe flare temperatures as a low as 4,000$K$, depending on the type of data products used for the analysis (see \autoref{sec:recommend}). 

\begin{figure}[ht!]
    \centering
    \includegraphics[width=0.48\textwidth]{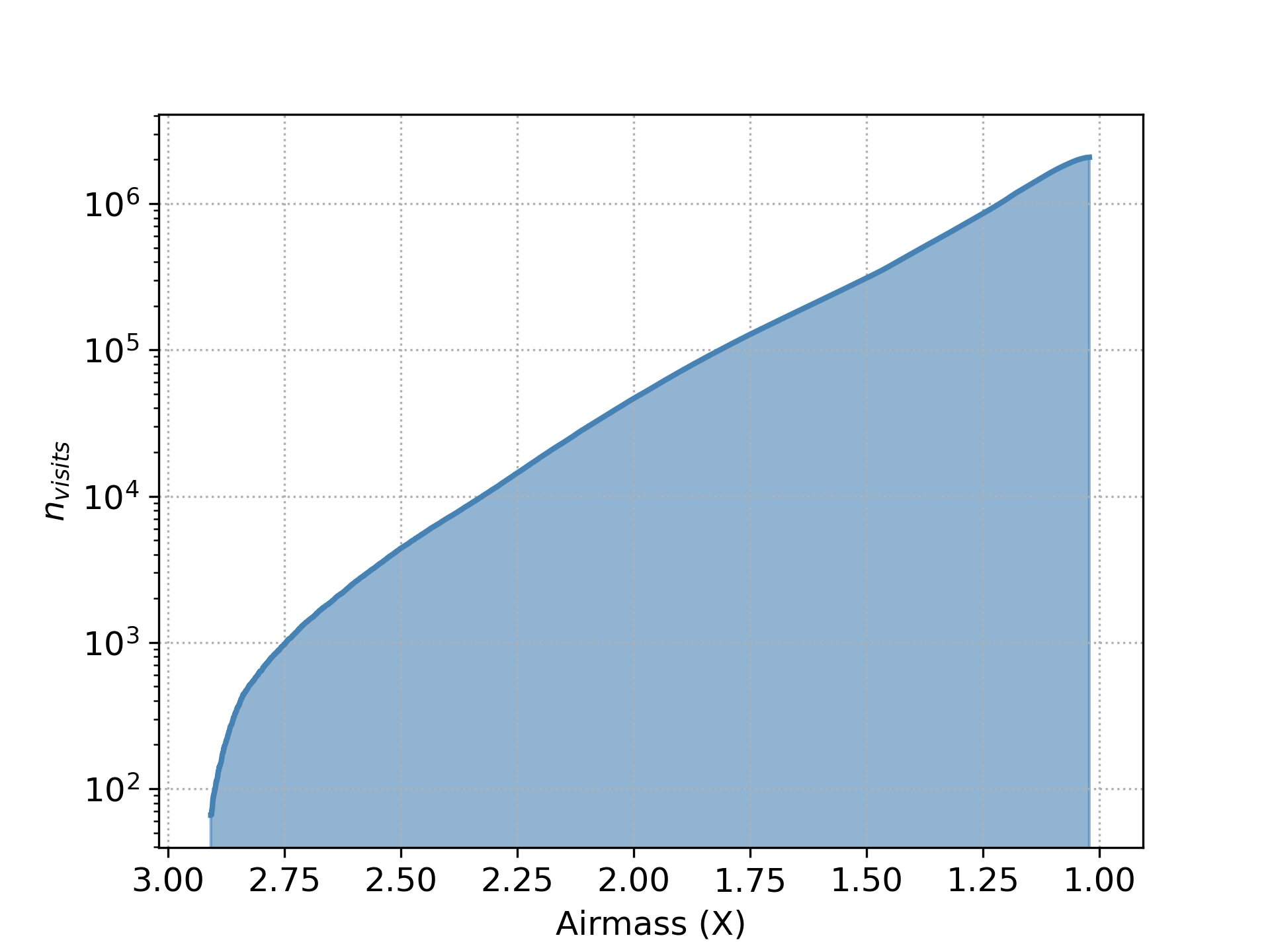}
    \caption{Airmass distribution of the current LSST survey strategy proposal (\texttt{baseline\_v3.0\_10yrs}, \citealt{PSTN-055}). Cumulative distribution showing the number of visits \textit{at or below} a given airmass. Plot produced within the Rubin Metric Analysis Framework \citep{maf}.}
    \label{fig:airmasshist}
\end{figure}

To maximize image quality, the automated LSST scheduling system will preferentially observe at low airmass by design \citep{LPM-17}. The dependence of this effect on airmass then begs the question of whether or not LSST will perform observations at sufficiently high airmass for the effect to be detectable. However, the most recent fiducial simulation of the LSST  (\texttt{baseline\_v3.0\_10yrs}, \citealt{PSTN-055}) suggests that nearly $10^6$ WFD visits will be performed above airmass 1.25 (\autoref{fig:airmasshist}), which, as we have shown in \autoref{sec:rubincapability}, is sufficient to produce a detectable \dDCR\ at typical flare temperatures (we will return to the LSST design in \autoref{sec:recommendopsim}). 

These simple calculations indicate that Rubin should be capable of detecting even relatively cool flares at typical airmass on a typical M5 dwarf. However, contamination from image warping, chromatic aberation, DIA misalignments, \emph{etc.} will complicate this simplified expectation.

\section{Precursor Surveys} \label{sec:precursor}

We tested our method on detected flares in precursor surveys selecting as flare candidates transient events with short duration ($\lesssim 2$-hour) and a change of magnitude of at least 0.4 mag in $g$ band.

\subsection{Zwicky Transient Facility}

The Zwicky Transient Facility (ZTF, \citealt{bellm2018}) was designed to detect transient objects across the entire northern hemisphere. It is considered a precursor survey for LSST and it will deliver images and lightcurves for 3,750 sqdeg/hour with alerts delivered in real-time and a significant fraction of its data made available without proprietary restrictions. The exposure time is similar to that of Rubin’s LSST (30 seconds + 10 seconds readout) to a single image depth of ZTFr~20.5, and three filters are available (ZTFg, ZTFr, ZTFi) also with similar throughput to the Rubin filters. With a smaller footprint by a factor of four and three filters the revisit time is nominally nearly one order of magnitude shorter than LSST's, enabling multiple observations of the same flare to be collected.

Notable differences, however, for the purpose of our science, are the overall data throughput, about 10 times smaller than Rubin as measured in bytes of data, but which corresponds to a factor of nearly 100 fewer targets due to the decreased limiting magnitude, and a significant decrease in image quality, measured as Point Spread Function (PSF), leading to decreased astrometric accuracy. The instrumental pixel scale is 1.01 arcsec/pixel, compared to Rubin’s $\sim$0.2 arcsec/pixel, and the median seeing-limited PSF is $\sim$2 arcsec, compared to $\lesssim1$ arcsec expected for Rubin. This forced us to limit the study to flares observed at high airmass to enable DCR-based temperature estimates, further decreasing the size of our flare sample. In our preliminary analysis, we inspected a sample of 17,000 bright \dM\ in ZTF; 2.3\% showed possible large flares in data collected airmass above 1.4, for a total of 414 candidates. To increase the confidence that the observed brightening was indeed a flare we required more than one observation within the event (\ie\ within $\sim2$~hours). Of these 414 flares, only one was captured with more than one data point. ZTF's reconstructed astrometric accuracy per science image with respect to
Gaia DR1 is $\sim0.045-0.085''$ for sources extracted at a 10-$\sigma$ limit, however, the seeing Full Width Half Maximum (FHWM) in this pair of images was 3.474'' for the first flare epoch and 2.910'' for the second. The seeing turned out to be the bottleneck in the application of our method; the astrometric solution generated by the ZTF pipeline measured a small $\sim0.2''$ displacement between the first and second image compared to other stars in the field, too small compared to the seeing to confidently determine whether or not the star moved towards the zenith.

\subsection{DECam Deep Drilling Fields}\label{sec:DDF}

The Dark Energy Camera (DECam; \citealt{flaugher2015}) at the Blanco 4-meter telescope is the instrument that enabled the Dark Energy Survey \citep{dark2016}. The camera itself is a single chip of the same kind as those that will constitute the LSST camera mosaic, making DECam data naturally comparable with LSST's, with high image throughput, similar image quality (0.263 arcsecond/pixel resolution, with a telescope located near the site of LSST leading to similar sky properties), and similar system wavelength coverage ($grizY$ filters).

\begin{figure}[ht!]
    \centering
    \includegraphics[width=0.48\textwidth]{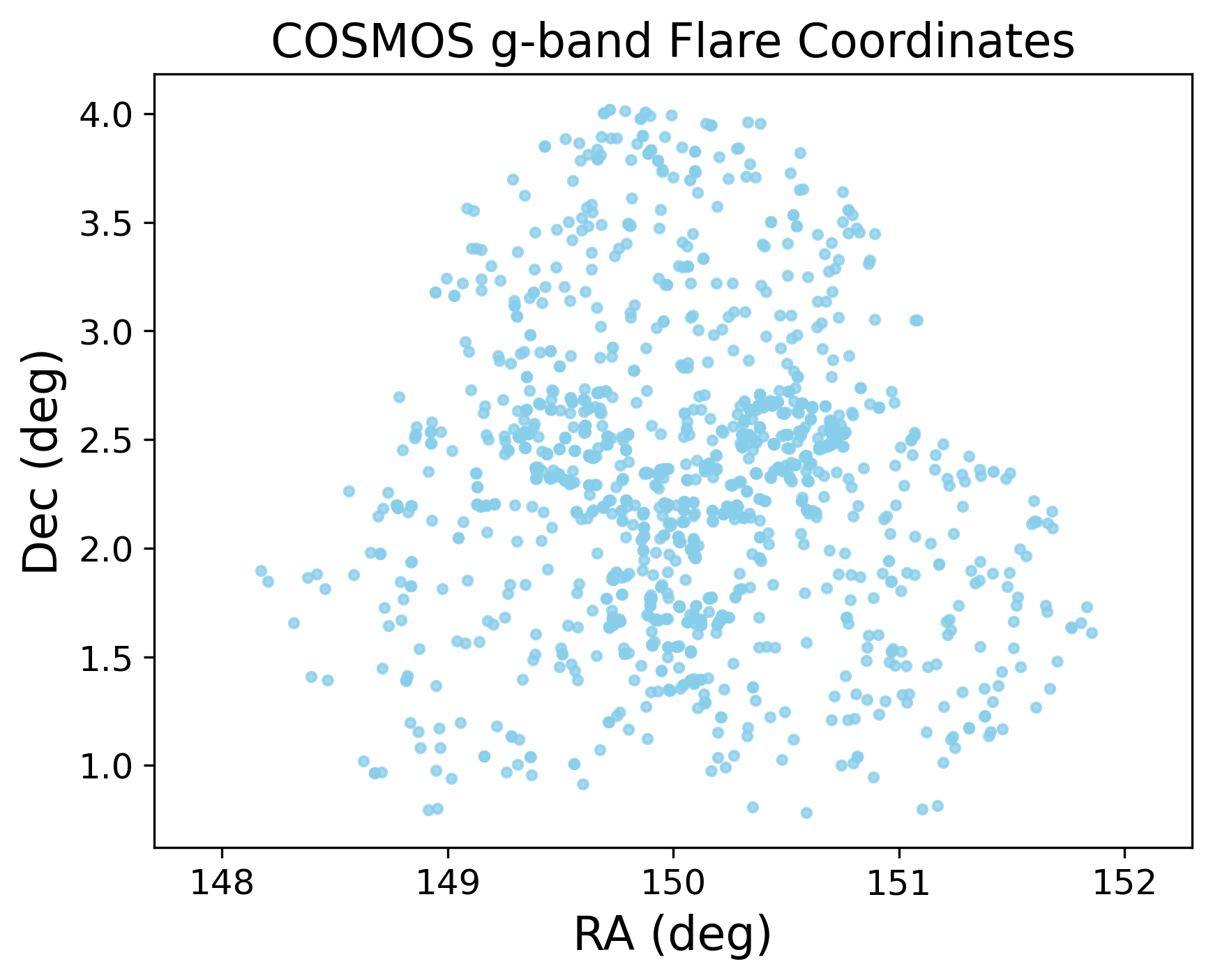}
    \caption{Location of DECam DDF flare candidate $g$-band objects in the COSMOS field.}
    \label{fig:decamcoords}
\end{figure}

\citet{graham2023} used DECam to survey two of the LSST DDFs: COSMOS \citep{scoville2007} and ELAIS \citep{oliver2000}. This led to a precursor survey of the LSST DDFs with 5-sigma
limiting magnitudes $r\sim23.5$~mag (single exposure). 
However, in addition to the shallower depth, there are other significant differences. This DECam DDF program was one of the co-founding members of the DECam Alliance for Transients (DECAT), within which multiple PIs of DECam programs pooled their time to enable dynamic queue-like scheduling and time-domain science. 
With a field of view of 9.6 deg$^2$, the LSST DOE camera will cover each field with a single pointing. Conversely, the DECam field of view (3 deg$^2$) is smaller and DECam DDF program used three adjacent pointings in COSMOS and two in ELAIS. Every night of observations cycled through each pointing five times, obtaining a sequence of $gri$ images per pointing. However, because the fields are covered with multiple pointings, 10 or 15 observations per night are obtained in the area where the  pointings overlap (see \autoref{fig:decamcoords}).  In contrast, LSST will obtain tens of consecutive images in five filters on each night when a DDF will be observed. Taking these differences into consideration, the DECam DDF program is an effective test-bed for our DCR flare studies, but not for temporally resolved flare investigations.

\begin{figure*}[ht!]
    \centering
    \includegraphics[width=0.8\textwidth]{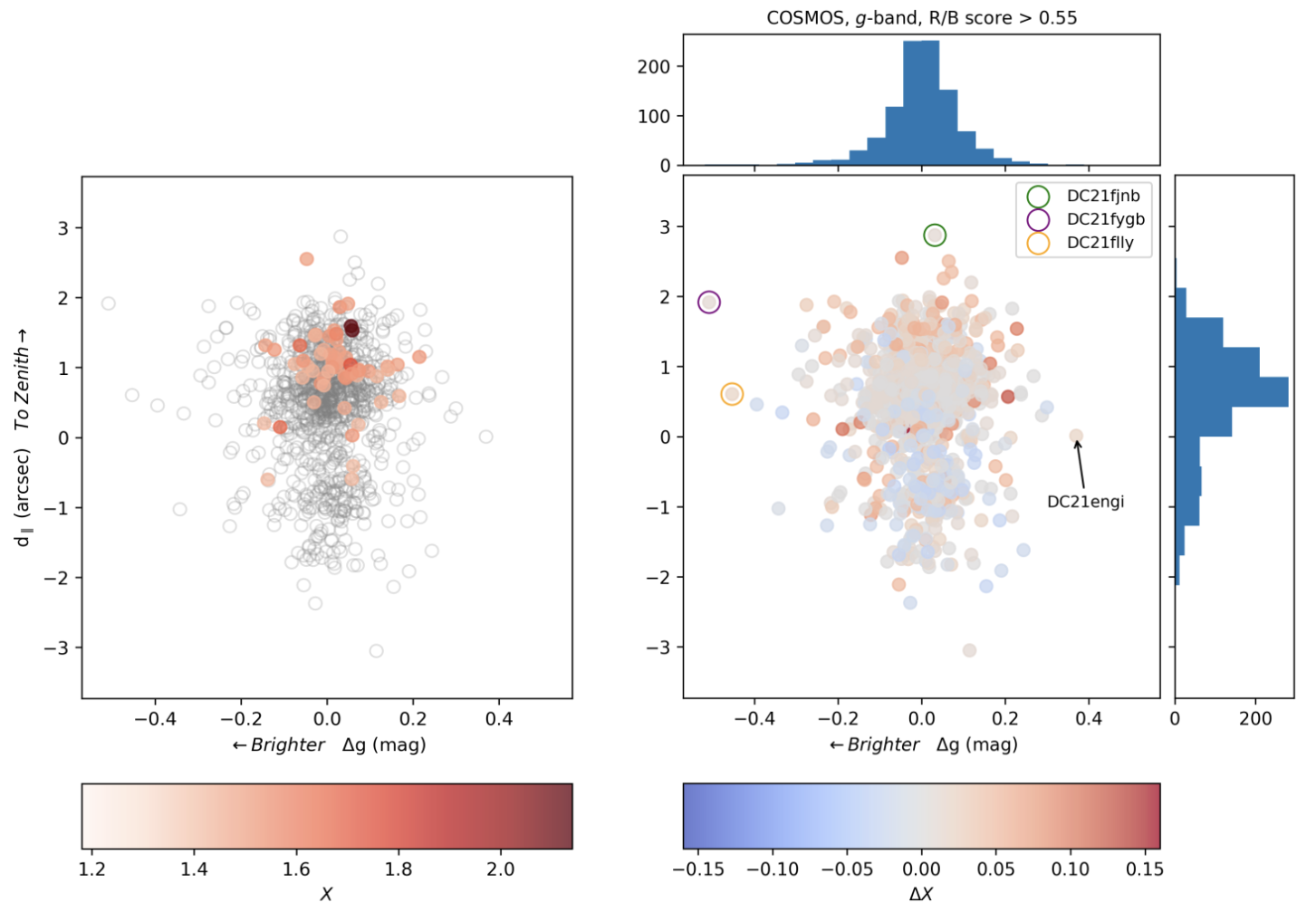}
    \caption{Scatter plots of 1015 flare candidates in the DECam DDF survey of the COSMOS field (for selection criteria  discussed in \autoref{sec:DDF}). The $x$ axis is the change in $g$ magnitude between the objects and the $y$-axis is the change in the component of the source's movement in the direction of the zenith. Each point represents one unique candidate ID with at least two DIA ``objects'' (two distinct observations). The change in magnitude and position are calculated as the difference between the two objects with the largest absolute change in magnitude.  \textit{Left panel:} For observations with airmass $X < 1.2$, points are shown as grey circles. Points are colored by the airmass of the initial detection for airmasses $X > 1.2$.  \textit{Right panel:} Each point is colored by the change in airmass between the two objects and histograms show the marginalized distributions of $g$ magnitude change (\emph{top}) and zenith-bound displacement (\emph{right}). Three candidates of interest, as described \autoref{sec:DDF}, are circled. The single candidate whose quiescent counterpart can be identified in the Gaia DR3 dataset, DC21engi, is marked by an arrow.}
    \label{fig:decamdmdd}
\end{figure*}
Importantly, the data processing pipelines for DECam DDF and LSST also differ: LSST will process its data with dedicated pipelines \citep{bosch2018}, while the DECam DDF fields data are processed with existing software to detect transient events \citep{graham2023}.
Briefly, the difference-image analysis (DIA) pipeline implemented in this survey ingests raw images directly from the NOIRLab data archive and performs standard data reduction procedures. 
\texttt{Source Extractor} \citep{bertin1996} is used to detect all sources in the image, \texttt{SCAMP} \citep{bertin2006} is used to calculate the astrometry for each chip and match each source with stars drawn from the Gaia DR2 catalog, and then \texttt{SWARP} \citep{bertin2010} is used to solve for the world coordinate system (WCS) using these objects. Object catalogs are generated from the reference images with \texttt{Source Extractor} and aligned with \texttt{SCAMP} and \texttt{SWARP}. The image subtraction is done with \text{HOTPANTS} \citep{becker2015}. Notably,
there is no DCR correction in the difference image analysis as implemented for the DECam DDF fields.
\texttt{Source Extractor} is used on the resultant difference image to identify residual signals and extract fluxes via forced photometry. Lastly, using the algorithm described in \citet{goldstein2015}, each detection is assigned a ``real/bogus'' score\footnote{
The real/bogus score is the output of a machine learning classifier designed to distinguish ``real'' sources from ``bogus'' detections (\ie\ image or subtraction artifacts and moving sources).}, ranging from 0 (bogus) to 1 (real). Objects (\ie\ detections) within 2'' of a previously detected object are associated with the same candidate ID.

We elected to focus our analysis on the COSMOS $g$ band sample (recall our preference for $g$ band from \autoref{sec:rubincapability}). We selected a subset of the publicly available DECam DDF transient catalogs\footnote{See Section 3.7.2 of \citep{graham2023} for access to these publicly available DECam DDF catalogs.} in the COSMOS field, requiring the following:
\begin{itemize}
    \item at least 1 g-band detection,
    \item all detections within 0.5 days (single-night), 
    \item a mean ``real/bogus'' score $\overline{RB}\geq 0.6$ across all detections. 
\end{itemize}
These cuts resulted in a sample of 1230 candidates, and the coordinates of this $g$ band detections' sample are shown in \autoref{fig:decamcoords}. 
To identify flares in the data and measure the change in magnitude during the event we applied the following additional cut, resulting in a final sample of 1015 COSMOS flare candidates: 
\begin{itemize}
    \item 
    at least two $g$-band detections, each with real/bogus score $RB \geq$ 0.55
\end{itemize}

We calculated $d_{\parallel}$, the component of the candidates' motion along the parallactic angle (toward or away from the zenith), for all pairs of $g$-band detections with the same on-sky association. $d_{\parallel}$ is plotted against the measured change in $g$ magnitude in \autoref{fig:decamdmdd} in order to visualize the sample in an informative phase space and select interesting candidates for additional inspection. We expected flares to be found in the upper left quadrant of the plot, \ie\ pairs of detections where the brightness of the source increased (relative to its initial difference image detection) and also moved on the sky in the direction of the zenith at the time of observation, or lower right quadrant for dimming flares moving away from zenith. In addition, we favor candidates observed at higher airmasses, but with little to no \textit{change} in airmass between observations, as a change in airmass dominates the chromatic contribution to DCR (see \autoref{fig:dcr_diagram2}).  However, as shown in the left panel of \autoref{fig:decamdmdd}, few observations of the DECam DDF fields occur at airmass above $X=1.2$, and none of these high-airmass detections displayed sufficient change in brightness to be considered for additional inspection.

\begin{figure}[ht!]
    \centering
    \includegraphics[width=0.45\textwidth]{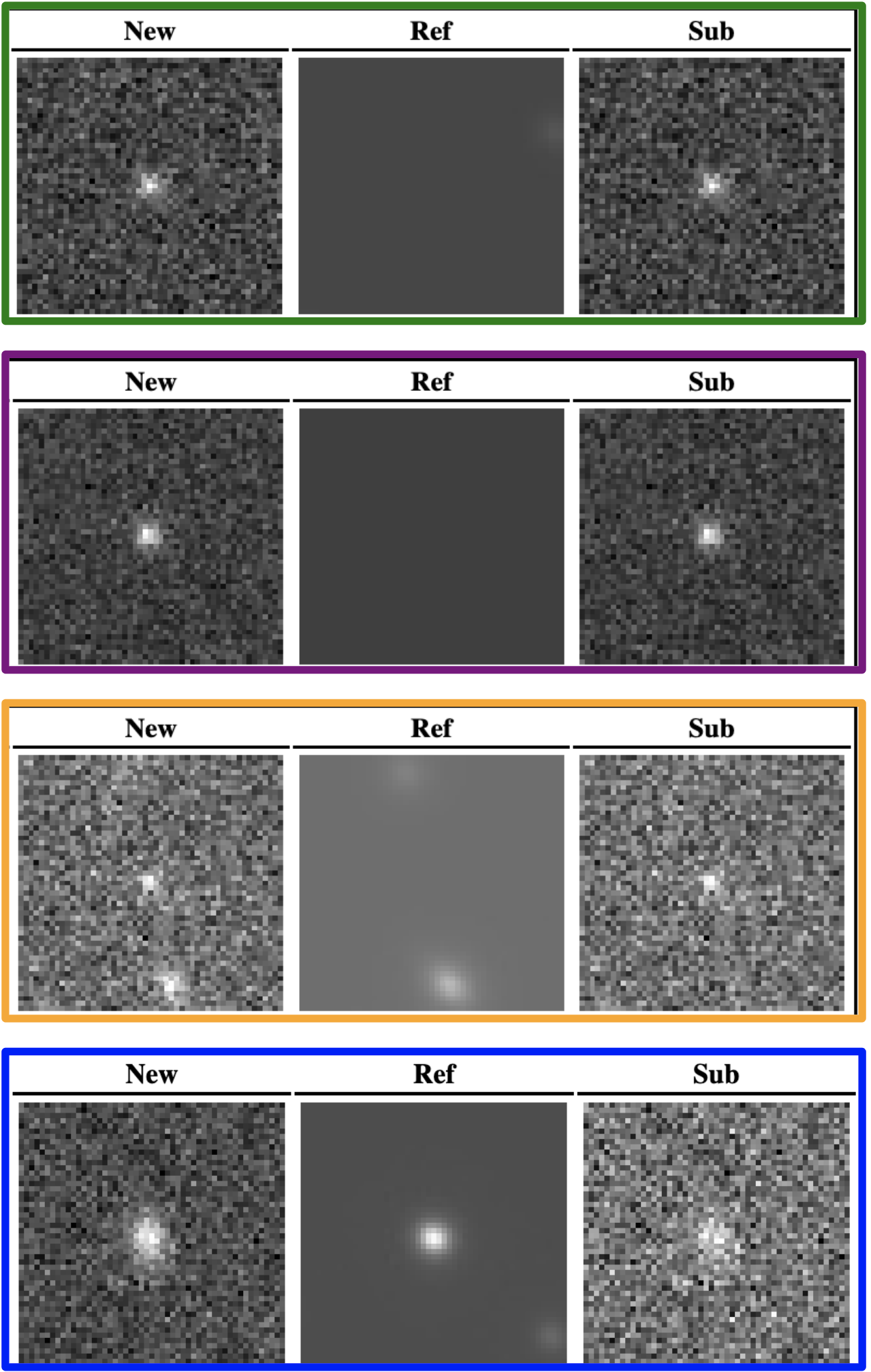}
    \caption{Image triplets for DECam DDF candidates, as shown by the DECAT LBL Pipeline Candidate Viewer. From top to bottom, the candidates are: DC21fjnb, DC21fygb, DC21flly,  and DC21engi. ``New'' denotes the search image, ``Ref'' the reference image, and "Sub" the difference image. A quiescent source consistent with a \dM\ is seen in the reference image for DC21engi.}
    \label{fig:triplets}
\end{figure}

\begin{figure}[ht!]
    \centering
    \includegraphics[width=0.45\textwidth]{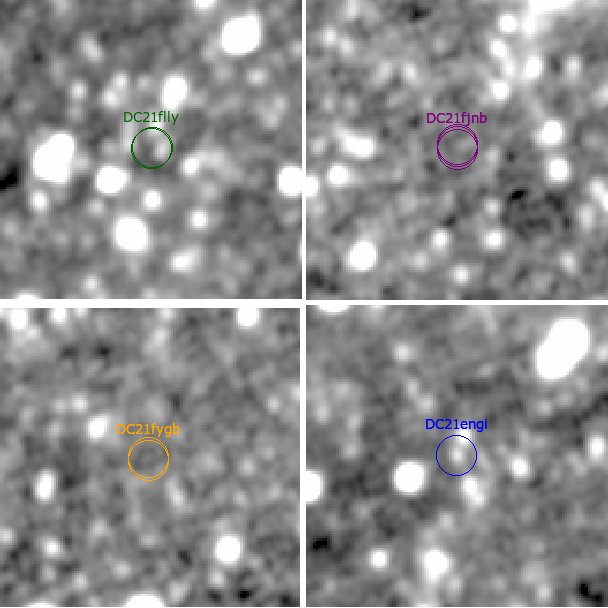}
    \caption{WISE W1 band images; the coordinates of candidate flares DC21flly, DC21fjnb, DC21fygb, and DC21engi are marked as labeled. In all panels, North is up and East is left. Each panel is 3.8' on the side. A 12" radius circle centered on each detection reported in Table \ref{tab:candidates} is plotted (where only one circle is visible, as for DC21engi, the detections are overlapping). No source can unambiguously be identified at the location of the transients, except for DC21engi. }
    \label{fig:wise}
\end{figure}
We selected three objects for further investigation (marked with circles in \autoref{fig:decamdmdd}): the two most extreme outliers in the upper-left quadrant that are ostensible rising flare candidates, and an object with large zenith-bound motion but no magnitude change, as a control. All of them show a small change in airmass between the two observations in a pair, as desired. While a few DECam DDF transients that passed our cuts are detected at high airmass $X>2$ (\autoref{fig:decamdmdd}, left panel), none of them have significant enough magnitude changes to be considered as flare candidates. All of our three objects of interest are observed at airmass $X\sim1.2$. Thus, in order to show a significant \dDCR, if they were flares, these first two candidate events would have to be extremely hot.
The coordinates, detection times, $g$ band photometry, and  $d_{\parallel}$ for these candidates are shown in Table \ref{tab:candidates}. To ascertain the stellar origin of these transients we look for a quiescent-state source in the templates.\footnote{We used Robert Knop's DECAT LBL Pipeline Viewer tool to search by Candidate ID and obtain the 51x51 pixel cutouts of the image triplets used in the DIA process.} The DIA triplets of the three candidates circled in \autoref{fig:decamdmdd} are shown in \autoref{fig:triplets}. From left to right, each triplet contains the ``new'' image (also called the ``direct'' or ``science'' image, obtained by the DECam DDF program), the ``reference'' (a template built from archival DECam images obtained in previous years), and the ``difference'' image (PSF-matched subtraction of the reference image from the new image). For none of the three candidates were we able to find a precursor source in the reference image. It is possible that the quiescent sources were not present in the reference image because they were beyond the survey magnitude limit. However, the templates reach about one magnitude deeper than the search images \citep{graham2023}, and a flare with 1\% surface coverage on an M3 dwarf,  would generate a brightening of $\sim$0.62 magnitudes compared to the quiescent source in $g$ band \citep{davenport2012}, which should enable the detection of the quiescent source even for transients detected close to the single image $5-\sigma$ limit (see Table \ref{tab:candidates}). Since \dM\ are typically bright in the infrared (IR) wavelengths, rather than the optical wavelengths, we also searched for a quiescent source in the IR by examining the WISE data \citep{wright2010} at the candidate's coordinates (\autoref{fig:wise}), but likewise did not see a source in these images. This is not surprising: given the effective magnitude limit of WISE W1 (16.6 mag; \citealt{cutri2012}), a faint transient in the DECam DDF data, such as our three candidates, is expected to be at the detection limit of the WISE data in quiescent state. This raises the possibility that the candidates may not be stellar sources, but rather Solar System Objects or extragalactic transients. We checked additional DECam alerts to see if the transients were detected at later epochs: extragalactic transients have typically slower evolutions, so they should remain detectable in successive nights. Inspecting data with the DECAT LBL Viewer, we found no subsequent alerts detections for any of the four candidates. Finally, we  checked if any known Solar System Object (SSO) is expected at the coordinates of the detection by inspecting the IAU Minor Planet Center (MPC) catalogs using the Minor Planet Checker\footnote{https://minorplanetcenter.net/cgi-bin/checkmp.cgi} and we found SSOs within one arcminute (the minimum search radius in the MPC) of the candidate coordinates, at the time of detection for all three candidates. This leaves asteroids as the most likely source of contamination given the brightness of the transient and the fact that each only appeared in a single night of observations.

\begin{deluxetable*}{llllcc}\label{tab:candidates}
\tablecaption{Candidate ID, Coordinates, timestamps, and $g$ magnitudes with errors, as reported by the DIA pipeline described in \autoref{sec:DDF}, for each DIA detection of each candidate circled in \autoref{fig:decamdmdd} and DC21engi. The coordinates, detection time, magnitude, and component of the motion in the direction of zenith as measured between the first and each successive observation are indicated.}
\tablehead{
\colhead{Candidate ID} & \colhead{RA} & \colhead{Dec} & \colhead{Object Datetime} & \colhead{$g_{DIA}$} & \colhead{$\delta g_{DIA}$}\\
\colhead{} & \colhead{($^{\circ}$)} & \colhead{($^{\circ}$)} & \colhead{(UTC)} & \colhead{(mag)} & \colhead{(mag)}
}
\startdata
DC21flly & 150.114428 & 2.698361 & 2021-04-15 01:08:45.598 & 23.143 & 0.110\\   
DC21flly & 150.114194 & 2.698476 & 2021-04-15 01:45:47.624 & 22.690 & 0.082\\   
DC21fjnb & 150.384041 & 2.705729 & 2021-04-15 01:08:45.598 & 22.627 & 0.065\\   
DC21fjnb & 150.384017 & 2.706228 & 2021-04-15 01:20:48.720 & 22.610 & 0.076\\   
DC21fjnb & 150.383980 & 2.706525 & 2021-04-15 01:27:32.945 & 22.659 & 0.067\\   
DC21fygb & 150.531830 & 3.533134 & 2021-04-18 00:43:48.466 & 21.833 & 0.041\\   
DC21fygb & 150.531737 & 3.533659 & 2021-04-18 01:35:23.940 & 21.325 & 0.026\\
DC21engi & 151.678533 & 2.169509 & 2021-04-09 02:23:03.576 & 22.617 & 0.111\\
DC21engi & 151.678564 & 2.169518 & 2021-04-09 02:41:18.079 & 22.987 & 0.115\\
\enddata\label{tab:candidates}
\end{deluxetable*}
\begin{figure}[ht!]
    \centering
    \includegraphics[width=0.48\textwidth]{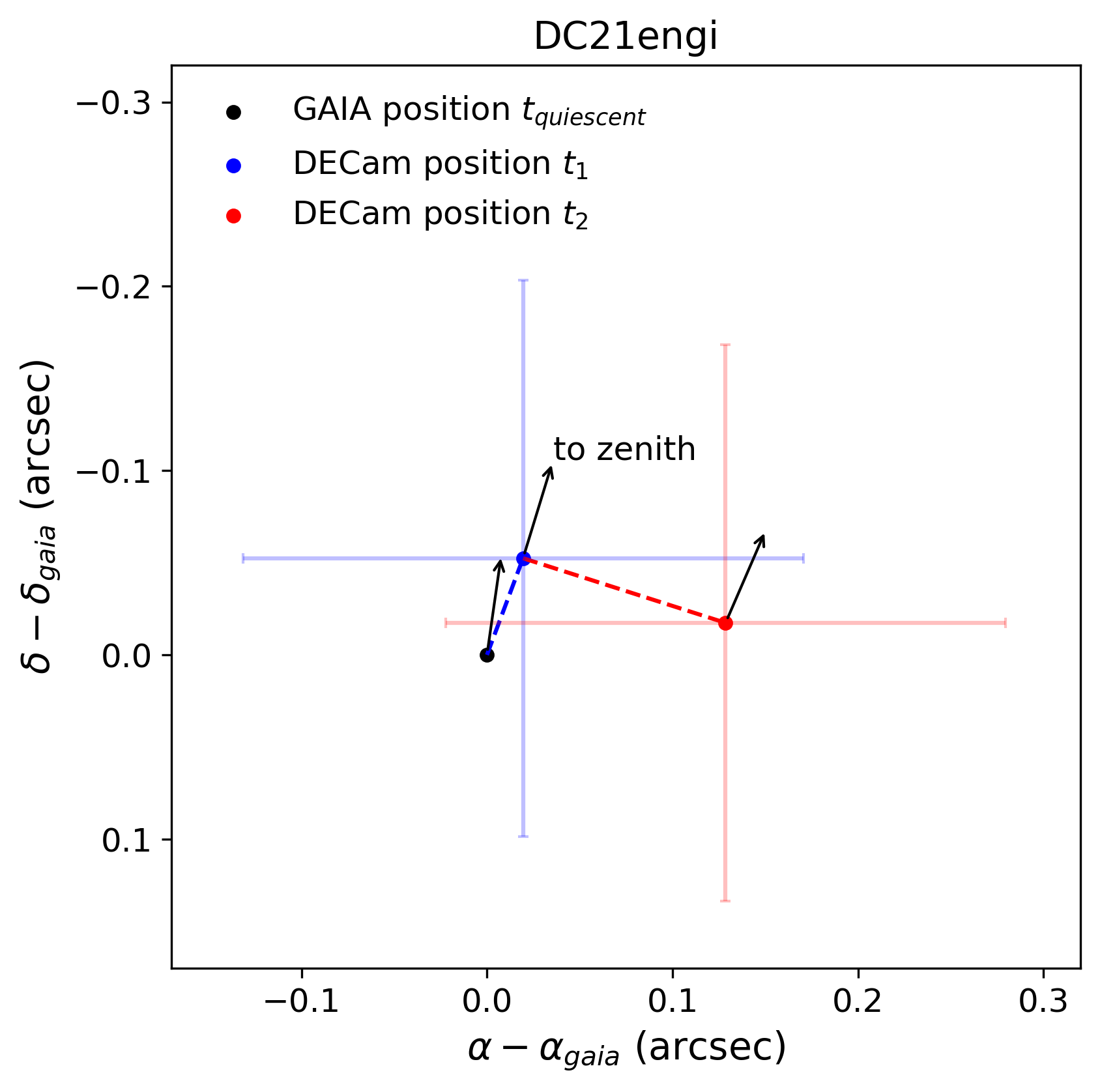}
    \caption{Schematic of the position change of candidate DC21engi during a DECam transient event captured in two detections, at times $t_1$ (2021-04-09 02:23:03.576) and $t_2$ (2021-04-09 02:41:18.079). The Gaia DR3 coordinates of the source are shown in black, and the coordinate axes are shifted such that the Gaia position is at the origin. The source is assumed to be quiescent at $t_{quiescent}$ (2021-04-09 02:00:00), and the direction to zenith at $t_{quiescent}$ and the position of the source in the DECam DIA subtraction are shown in blue. The maximum astrometric residuals (0.15'') are indicated by the error bars on the DECam event positions. Arrows point to the zenith-bound direction at the time of the observation, showing the axis along which the next observation should be found if its motion were dominated by DCR.}
    \label{fig:dc21engi}
\end{figure}

\begin{figure*}[ht!]
    \centering
    \includegraphics[width=0.8\textwidth]{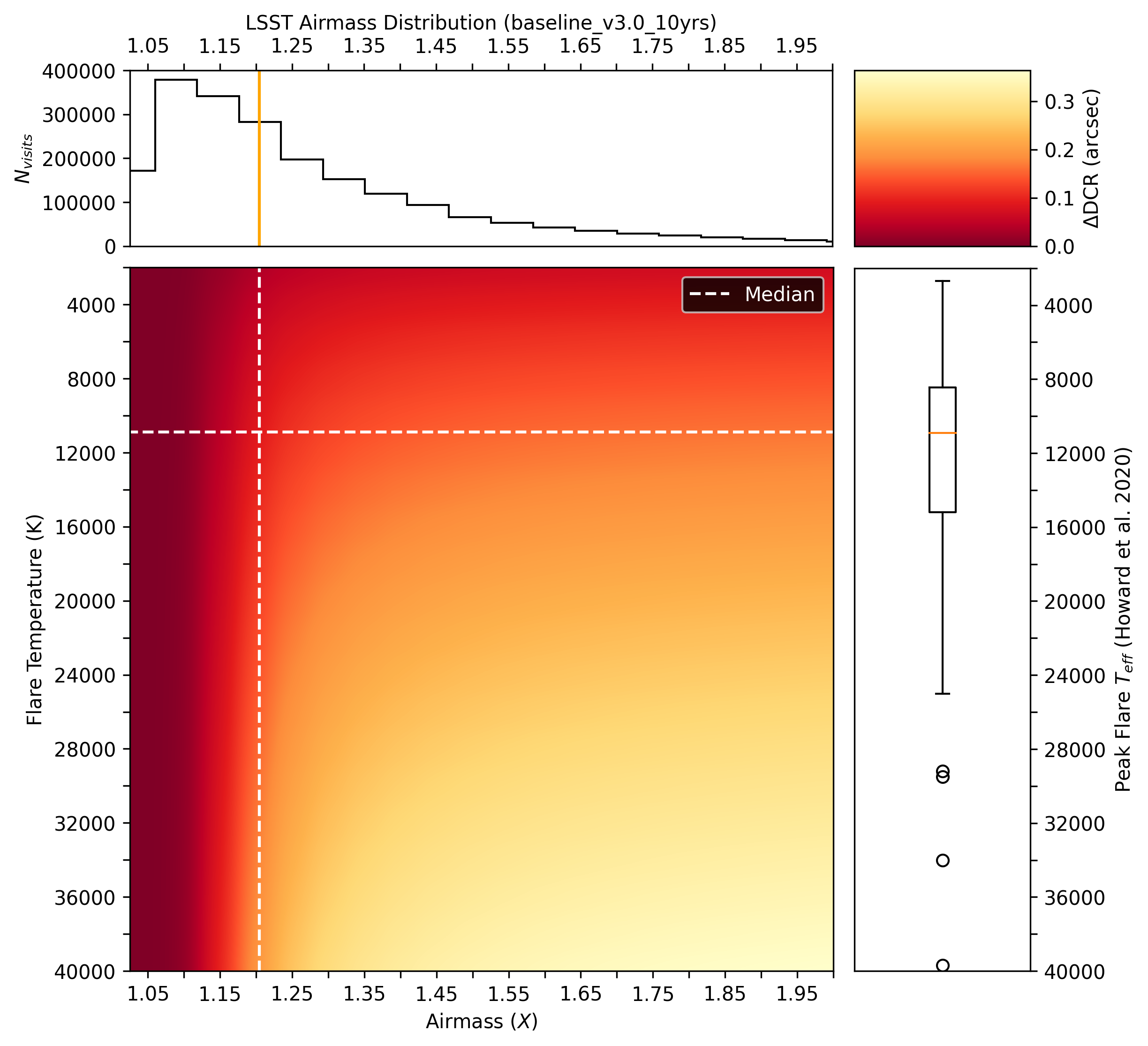}
    \caption{\textbf{Center:} \dDCR\ induced as a function of flare temperature and airmass in LSST $g$-band. \textbf{Right:} Box-and-whiskers plot of peak effective flare temperatures measured by \cite{howard2020}. \textbf{Top:} histogram of the per-visit airmass in the current LSST baseline observing strategy (\texttt{baseline\_v3.0\_10yrs}, \citealt{PSTN-055}). The median airmass and temperature are indicated by orange lines.}
    \label{fig:tempam}
\end{figure*}

\begin{figure}[ht!]
    \centering
    \includegraphics[width=0.48\textwidth]{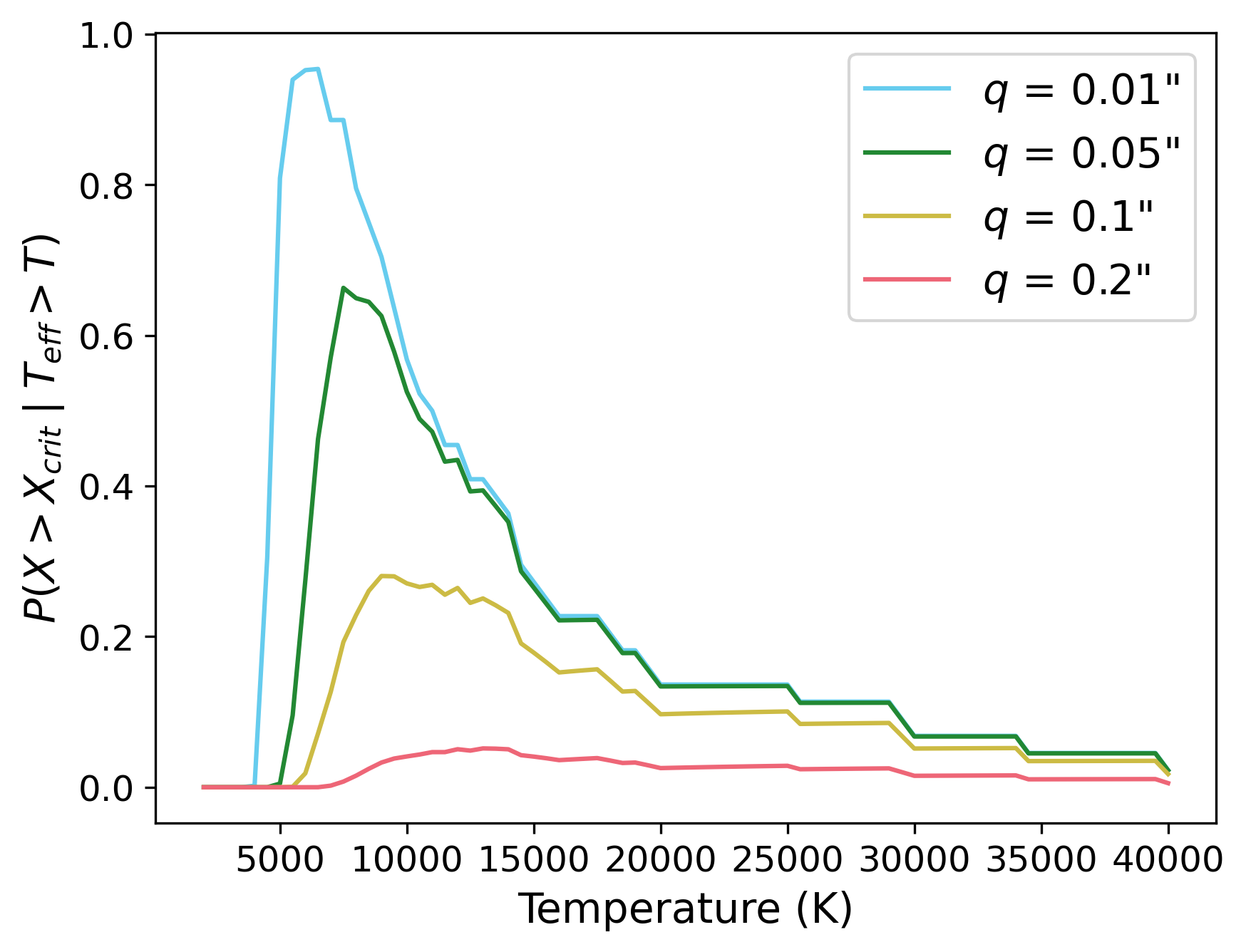}
    \caption{Probability of detecting \dDCR\ induced by a flare at or above peak effective temperature T for four different astrometric accuracy limits. The probability of DCR detection is measured as the conditional probability $P(X > X_{crit} | T_{eff} > T)$, where $X_{crit}(T,q)$ is the minimum airmass necessary for a flare of temperature T to produce a \dDCR\ greater than the astrometric accuracy $q$. The temperature distribution follows the measured flare temperatures in \citet{howard2020}. The small number of flares at large $T$ causes the discontinuities in the probability at $T_{eff} > 20,000K$.}
    \label{fig:dcrprob}
\end{figure}

\begin{figure*}[ht!]
    \centering
    \includegraphics[width=1\textwidth]{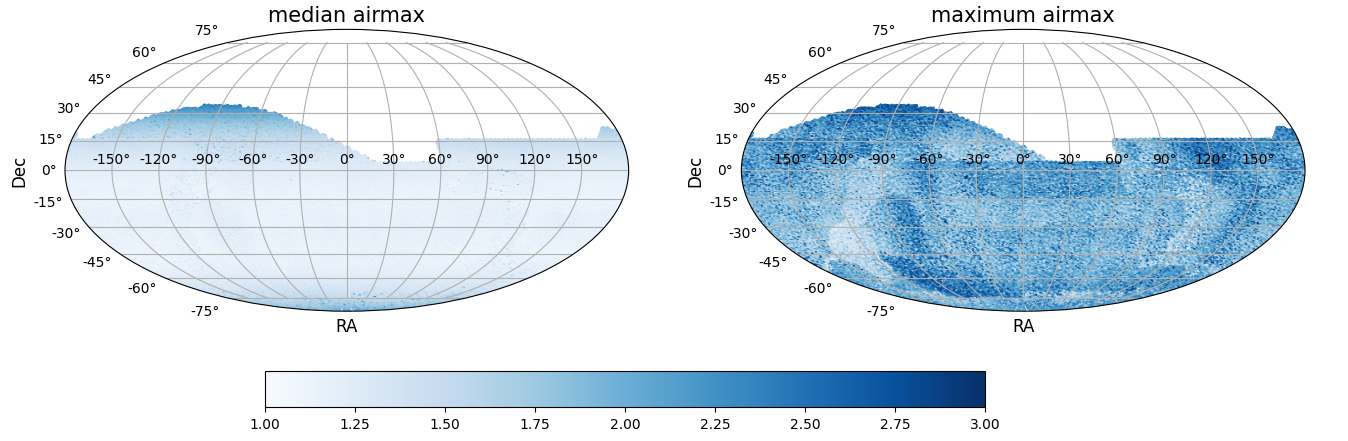}
    \caption{Airmass sky distribution of the current LSST survey strategy proposal (\texttt{baseline\_v3.0\_10yrs}, \citealt{PSTN-055}):  skymaps are shown for the median airmass (\emph{left}) and maximum airmass (\emph{right}) in $g$-band. Plots produced within the Rubin Metric Analysis Framework. \citep{maf}. To produce these maps, a sky segmentation into healpixels with resolution 64 \citep{healpix2005}}
    \label{fig:skymap}
\end{figure*}

\begin{figure}[ht!]
    \centering
    \includegraphics[width=0.48\textwidth]{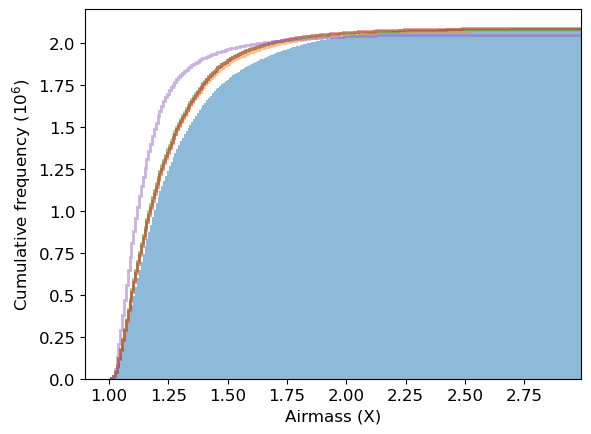}
    \includegraphics[width=0.48\textwidth]{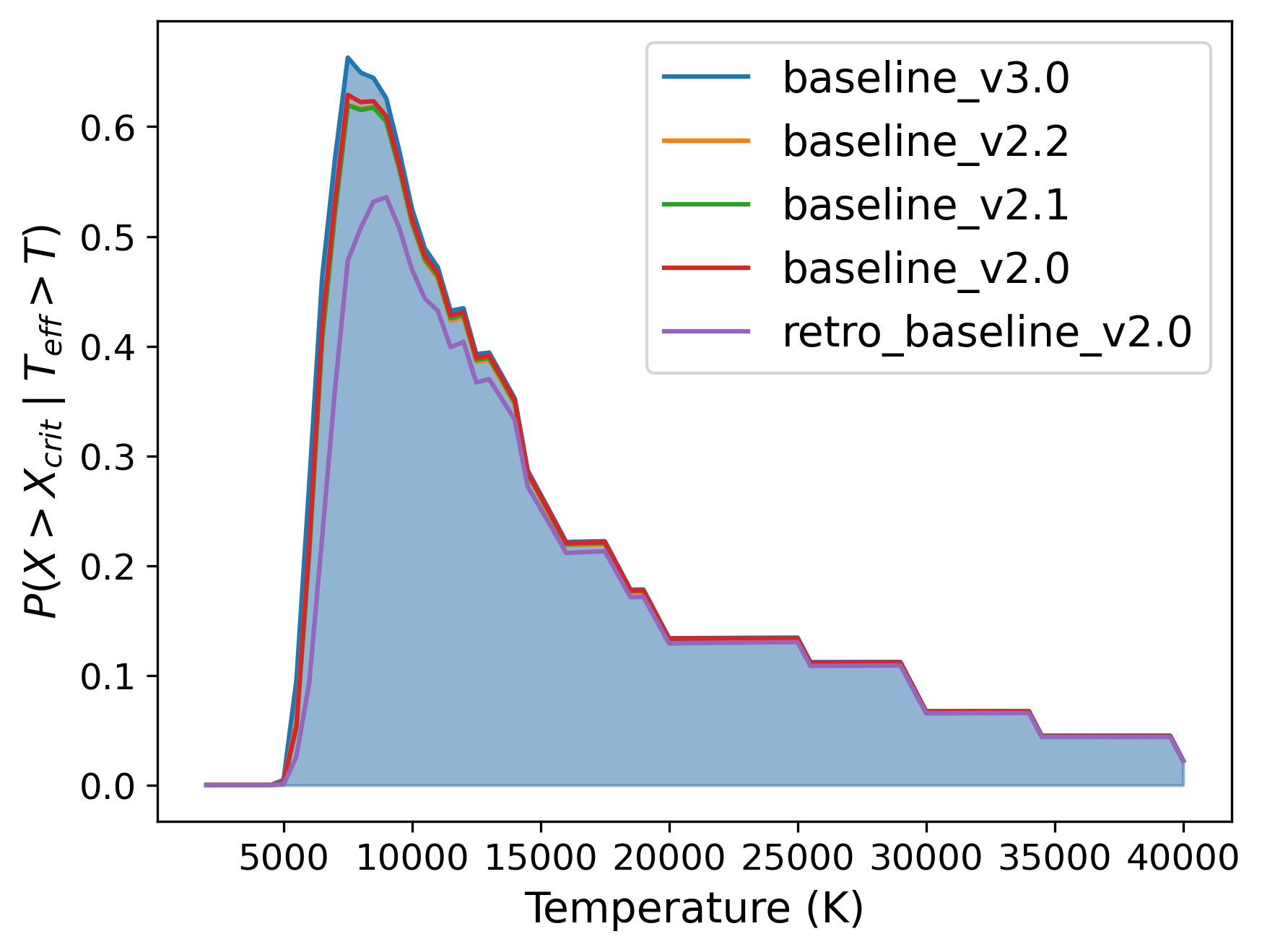}
    \caption{\emph{Top:} Distribution of observation airmasses over five LSST recent observing strategy proposals (as labeled). While it remains small, the planned fraction of high-airmass observations in LSST has grown over time. \emph{Bottom}: the corresponding probability of measuring a \dDCR: as \autoref{fig:dcrprob} but fixing $q=0.05$ and for the same five simulations of the LSST as in the top panel.}
    \label{fig:opsims}
\end{figure}

We then took a more traditional approach and began with ascertaining if any of the DECam DDF detections selected by our cuts were indeed flares, then searched for their location in the $d_{||}$-\emph{vs}-$\Delta g$ and see if we can detect a \dDCR. To do this, we crossmatched the DECam DDF source location with a sample of known cool stars in the COSMOS field from the Gaia DR3 catalog \citep{brown2021}. We selected Gaia sources with $G_{bp}-G_{rp} > 2$ and $M_G > 5$ to choose $K-M$ dwarfs and performed an additional cut on parallax error ($\leq 1''$) over three areas corresponding to the three DECam pointings that compose the DDF's COSMOS field. This led to the selection of 6993 Gaia DR3 sources. We then crossmatched the positions of the Gaia sources to within 1.0'' of the initial 1230 DECam DDF candidate positions using the \texttt{match\_to\_catalog\_3D} method of the \texttt{astropy.coordinates.SkyCoord}
object. This crossmatch left us with a single candidate, DC21engi, whose DIA triplet for two $g$-band detections is shown in \autoref{fig:triplets} and whose WISE W1 image is shown in \autoref{fig:wise}.

We analyzed the astrometric changes of DC21engi between quiescence and event by taking the Gaia coordinates as the bonafide quiescent position. In \autoref{fig:dc21engi} we examine the relative positions of the first and second DIA $g$ band detections, using the parallactic angle to indicate what the motion of the source would be if it moved directly toward zenith as expected for DCR. While the initial detection ($g\sim22.6$) of the event and the associated motion from its Gaia position is suggestive of DCR (\ie\, in the zenith direction), the second detection ($g\sim23.0$) defies the expectation that the source would move back towards its quiescent location as the photosphere cools and the object position is measured in a direction perpendicular to the parallactic angle. The bottleneck here is likely the precision of the DECam astrometric solution, which dominates the uncertainty in the position: astrometric match requirements for images in \citet{graham2023} are set by median astrometric residual across the entire field which must not exceed 0.15''. This margin is shown by the errorbars on the detection positions, indicating that this candidate does not conclusively demonstrate motion (produced by DCR or otherwise) during the transient event.

\section{Technical Recommendations for \dDCR\ Detection}\label{sec:recommend}

The difficulties encountered in the process of validating this method on precursor surveys emphasize the extremely high image quality and astrometric fidelity requirements demanded by such a method. We propose a set of astrometric accuracy benchmarks to maximize atmosphere-aided studies, continuing to use flares as our case study. \autoref{fig:tempam} shows the \dDCR\ induced as a function of flare temperature and airmass in LSST $g$-band, assuming a distribution of flare $T_{eff}$ as measured by \citet{howard2020} using a combination of \textit{Evryscope} \citep{law2015} and \textit{TESS} data, modeling flares as blackbodies, and a per-visit airmass as in the \texttt{baselinev3.0\_10yrs} simulation of LSST.

\subsection{Astrometric requirements}. 
By combining these distributions, we can estimate the probability of measuring \dDCR\ in LSST as a function of flare temperature. \autoref{fig:dcrprob} shows the conditional probability of producing a \dDCR-induced astrometric shift detectable in LSST during a flare given the flare temperature distribution in \citet{howard2020}: $P(X > X_{crit} | T_{eff} > T)$, where $X_{crit}(T,q)$ is the minimum airmass necessary for a flare of temperature T to produce a \dDCR\ greater than the astrometric accuracy $q$. The probability of measuring \dDCR\ from a flare with peak temperature at least 10,000$K$ is 56.8\% for a survey with astrometry accurate to within 10 mas, but this probability falls to 27.1\% at the 100 mas level and 4.1\% at the 200 mas level.

New methodologies for increasing astrometric accuracy have been proposed that could improve upon the 100 mas requirement set in \citet{LPM-17}. \citet{fortino2021} leveraged Gaia astrometry on stars visible in both surveys and Gaussian process regression (GPR) to reduce the astrometric variance caused by atmospheric turbulence. Testing on the orbit of trans-Neptunian object Eris ($r\approx18.5$) they found that GPR corrections reduced the root-mean-square (RMS) residuals in $riz$ band from 10 to 5 mas. The observations used to perform the correction to the Eris orbit were done at median airmass $X=1.29$ which exceeds the median airmass of \texttt{baseline\_v3.0\_10yrs} setting the expectation that similar improvements could be observed in \dDCR\ relevant LSST observations. Furthermore, they state explicitly that the GPR technique should be applicable to LSST, and that Rubin’s larger aperture compared to DECam means that more stars will be available for the fit in the turbulence-dominated leading to improved results. We note that the RMS reduction is proportional to the number of Gaia stars available in the image, indicating that better results (up to a factor of 5) can be expected in Galactic fields, where the rate of flares is enhanced (see \citealt{fortino2021} fig. 5)

Full characterization of the uncertainty in both the quiescent and flaring position of the source is necessary for DCR-based inference of flare temperature. If prompt follow up is not required, or even not possible for very short duration events such as flares, the Rubin annual Data Releases can be used for the analysis, leveraging the high Rubin's internal relative astrometric accuracy (\autoref{fig:deltashifttemp}). This analysis can also be performed using the world-public alert packets sent to brokers within 60 seconds of observation and the Prompt data products released 24 hours after observation \citep{LSE-163}. However, these data products do not include positional uncertainties, thus direct analysis of the images is required.

\subsection{The impact of observing strategy choices}\label{sec:recommendopsim}
The observing strategy of Rubin LSST, which is still being finalized at the time of writing of this manuscript \citep{bianco2021}, has led to the current survey strategy recommendation \citep{PSTN-055}. The the survey strategy is simulated via the LSST Operation Simulator \citep{2014SPIE.9150E..15D, opsim, 2019AJ....157..151N}; the median and maximum airmass of the LSST current observing strategy proposal (\texttt{baseline\_v3.0\_10yrs}) is shown in \autoref{fig:skymap}. High-airmass visits are favorably distributed across the WFD footprint, with visits above airmass 2.0 occurring at least once anywhere in the WFD footprint. 

Four overarching science goals set the survey requirements and the high level design of the survey strategy \citep{ivezic2019}: probing dark energy and dark matter, building an unprecedentedly complete orbital catalog of Solar System objects, exploring the transient and variable sky, and advancing our understanding of the Milky Way via the resolved stellar population. The system is designed to simultaneously maximize field of view and depth, \ie\ to maximize the volume of space-time that can be surveyed, leading to a fast survey that can scan the entire southern sky in $\sim3$ days, which maximizes the discovery potential in time domain. Many of the science deliverables of LSST, however, require high-resolution imaging and exquisite image quality, which pushes the survey design to prefer low airmass observations. Nonetheless, in the complex optimization exercise that weighs in the needs of many science deliverables, it is inevitable that a subset of the observations will be performed at medium, and even high airmass.  Over the course of several years, as the observing strategy for LSST has evolved under the guidance of many community contributions \citep{bianco2021}, we observe that the power in the high-airmass tail of the distribution of airmasses has increased. As the optimization of the observing strategy is refined under a complex net of science goals, additional constraints on the pointing compete with optimizing the airmass choice for image quality. This tail of high airmass observations, however, is welcomed by our science case and all atmospheric-aided studies \citep{richards2018, yu2020} as it increases the probability of measuring ($\Delta$)DCR. \autoref{fig:opsims} shows the evolution of the distribution of airmasses over five recent \texttt{opsim}s (simulations of the LSST 10-year survey). The most recent recommendation for the Rubin LSST observing strategy \citep{PSTN-055} has led to the most favorable baseline among those considered in this work for DCR-assisted temperature and SED measurements.

\begin{figure*}[ht!]
    \centering
    \includegraphics[width=0.8\textwidth]{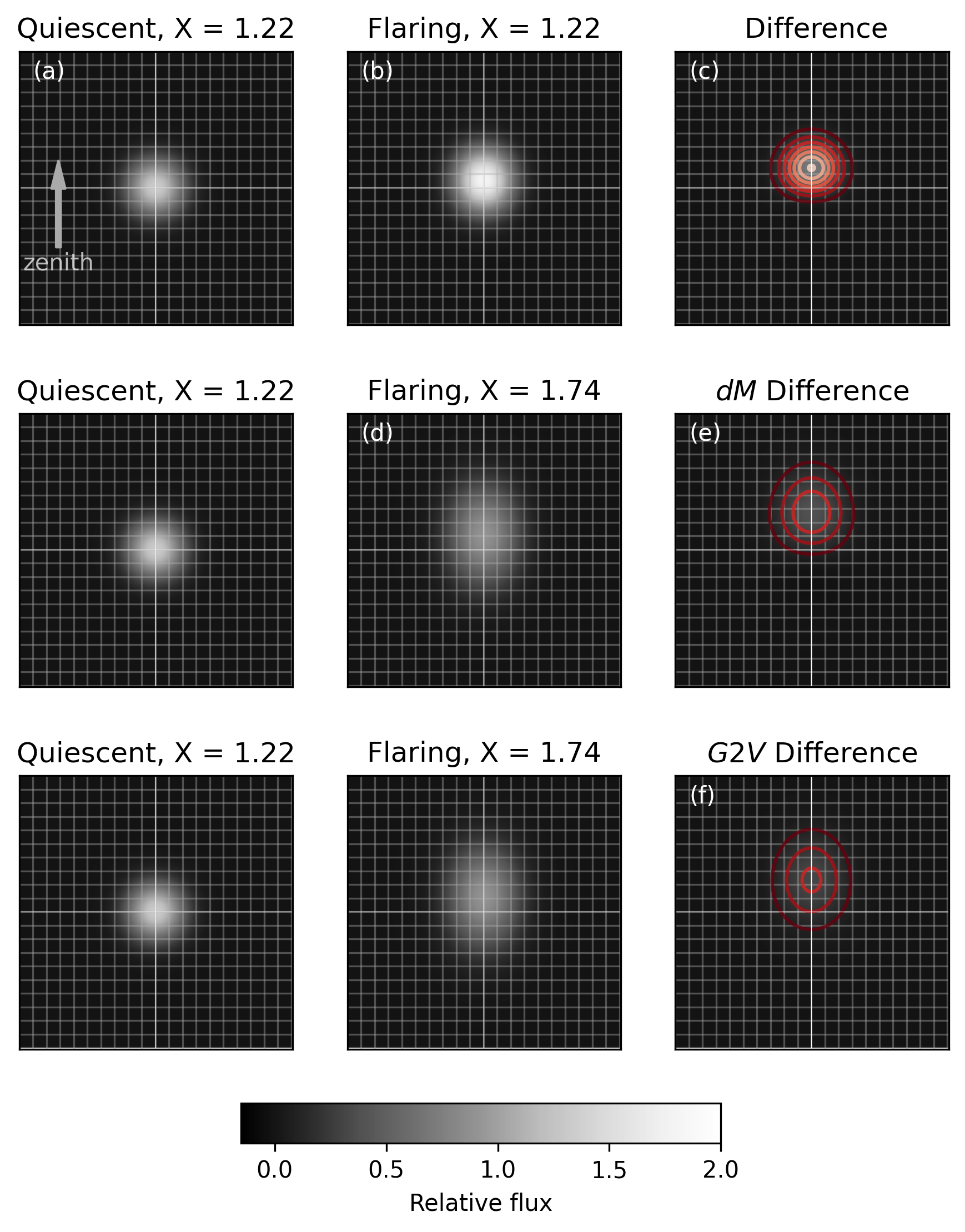}
    \caption{Simulation of the impact of three possible DCR correction strategies applied in $g$-band DIA. 
    The pixel scale of the LSST system (0.2 arcsec/pixel) is indicated by the grid and the zenith direction by an arrow. (\emph{a}): a quiescent \dM\ source observed at airmass $X=1.22$. 
    (\emph{b}): The source is observed during a $T = 10,000$ K flare at the same airmass. In both panels \emph{b} and \emph{d}, the flare flux is normalized such that the total flux at flare is 50\% greater than the total flux in quiescence. The difference of the flare and quiescent images is shown in the (\emph{c}); contours help the reader quantify the PSF width and centroid in the difference image. The quiescent component is subtracted cleanly in DIA and the residuals associated with the flare directly quantify \dDCR. Matching the DCR templates with correct assumptions on the quiescent SED would replicate this result. In 
    (\emph{d}), the same flare is observed at airmass $X=1.74$. DCR shifts the centroid of the quiescent source in the zenith direction by $\sim42$ arcsec. This shift is corrected in the visualization. The flaring PSF center is offset from the center of the image by $0.21$ arcsec due to \dDCR, and the PSF is significantly elongated in the zenith direction (with PSF variance increase $\Delta V =0.14$ arcsec$^2$). (\emph{e}): The quiescent and flare images are differenced after modeling the DCR effects on the quiescent source using the true \dM\ SED. The quiescent components is removed cleanly and the residuals can be used to measure \dDCR\ directly as in (\emph{c}). (\emph{f}): For the flare observed at airmass $X=1.74$ (\emph{d}), the same DCR correction as for (\emph{e}) are applied, but based on a G2V (solar type) quiescent SED: the flare and quiescent images are differenced resulting in non-zero residuals for the quiescent component and distorted residuals for the flare component. 
    }
    \label{fig:psfshift}
\end{figure*}

\subsection{Data Products Considerations}

The astrometric requirements of LSST are sufficiently demanding that DCR corrections 
may be required to achieve them (\citealt{LPM-17}, \citealt{LDM-151}), and understanding how and when these corrections are applied is necessary to develop a schema for measuring \dDCR\ with the Rubin data products. While the position of a flaring star will be available in an alert prompted by a flare, the centroid associated with the alert will be measured on the difference image. Difference imaging is at the core of Rubin's transient detection, and as Rubin is not equipped with an atmospheric dispersion corrector, DCR presents a challenge for image subtraction methods, most notably in the form of ``dipoles'' in the subtracted image caused by mis-subtraction of sources observed at different zenith angles. At the Difference Image Analysis (DIA) stage, DCR-matched templates can be used to mitigate these dipoles. This method (described in detail in \citealt{DMTN-037}) iteratively forward-models the unrefracted sky and uses the result as a template for image subtraction. However, in practice, LSST will not produce enough data in the bluer bands in the first year of operations to produce DCR-matched templates \citep{LDM-151}. 

The LSST Science Pipeline does not currently implement wavelength-dependent PSF modeling \citep[e.g.][]{meyers2015}
to account for DCR effects, although this will probably be added early in the survey.
DCR produces both a bulk shift of the PSF centroid (first order effect) as well as a second order effect on the shape of the PSF, increasing the dispersion along the zenith direction. Kolmogorov turbulence in the atmosphere leads to a linear isotropic contraction or dialation of the kernel.

In \autoref{fig:psfshift} we explore the impact on DCR correction approaches on the resulting DIA -- and thus our ability to infer flare properties -- for a simulated flaring M dwarf. The initial model (top row) shows the most ideal scenario, where the T=10,000K flare occurs at the same airmass as a template. The PSF is modeled as a Gaussian with $FWHM=0.7$ arcsec seeing. As expected, the resulting DIA shows a bulk shift toward zenith from the blue flare of $\sim0.10$ arcsec or $\sim 0.5$ pixels at the LSST system plate scale at an airmass of $X=1.22$. This model represents perfect DCR correction, since the template and flare occur at the same airmass, and would allow us to correctly infer the flare's temperature in an ideal (i.e. noise free) scenario.


The second scenario in \autoref{fig:psfshift} demonstrates the flare occurring at a higher airmass ($X=1.74$) than the quiescent template. If appropriate DCR correction for the red \dM\ SED is applied, then the resulting DIA is again due to the blue flare flux alone. Note that the DIA shift is larger and more broadened compared to the first scenario, due to the increased airmass, which naturally results in a more precise constraint on the flare temperature as expected (e.g. see \autoref{fig:deltagrid} and \autoref{fig:deltashifttemp}). This scenario represents the pipeline applying perfect DCR corrections for all known source SEDs at every airmass, making \dDCR\ measurable from DIA.


The third scenario shown in \autoref{fig:psfshift} demonstrates an incorrect or ad-hoc correction for DCR. Here we have assumed that the flare observation was at a high airmass, and the field was corrected for some average DCR effect without any knowledge of the quiescent source SED. For this demonstration, we assumed the \dM\ had a DCR correction applied for a solar-type G2V star. The impact here is an over-correction for DCR for the quiescent red \dM\ source. The blue flare flux still appears in the resulting DIA in \autoref{fig:psfshift}f. However, the resulting DIA signal is distorted from the expectation (\autoref{fig:psfshift}e) due to the incorrect DCR correction, which would negatively impact the accuracy of the inferred flare temperature. 

This result of these DCR simulations show that 1) the \dDCR\ signal should be a good indicator in DIA products of flare activity for low-mass stars at moderate airmasses, even with imperfect DCR corrections, and 2) careful and accurate DCR corrections based on the quiescent stellar SED are required to correctly infer flare temperatures using DIA.

\section{Conclusions} \label{sec:conclusions}

We have proposed a method to extract spectral information from chromatic transients by taking advantage of the change in differential chromatic refraction (DCR) between quiescence and event states, or \dDCR, using flares as a case study. We used a composite spectrum derived from SDSS \dM\ spectra to demonstrate the capability for LSST to measure the \dDCR\ induced by flares with effective blackbody temperature 10,000$K$ at or above $X$=1.2 in $g$ band, enabling \dDCR\ temperature measurement of flares over a significant fraction of total survey operations. Depending on the type of data products available for the analysis, flare temperatures as low as 4000$K$ could also probed by Rubin using DCR. In order to prepare to deploy this methodology on the Rubin pipeline, we tested its efficacy on two precursors to LSST, the Zwicky Transient Facility (ZTF, \citep{bellm2018}) and the ``Deep Drilling in the Time Domain with DECam'' program \citep{graham2023}. Our initial tests of the ZTF data quickly revealed a critical issue with using ZTF as a validation testbed, as the ZTF pixel scale (1.01''/pixel) is an order of magnitude larger than the astrometric change we expect to be induced by the \dDCR\ produced by a typical flare and the typical seeing exceeds 2''. While the system properties and observing conditions of the DECam DDF observations are more favorable, and indeed comparable to the LSST's, the processing pipeline did not lead to accurate enough astrometric solutions to enable \dDCR\ measurements. We estimated how accurate an astrometric solution must be in order to be able to measure \dDCR, given an observed flare's peak temperature and found that a survey must be accurate to within 0.01'' to measure 56.8\% of \dDCR\ events produced by flares above 10,000$K$.

This initial exploration of studying stellar flare temperatures using DCR in the context of Rubin LSST has demonstrated the potential for next-generation surveys to use incredible advances in image quality, astrometric solutions, and data throughput to break new ground using novel, albeit unconventional methods. However, the challenges encountered in our search for \dDCR\ in precursor surveys emphasize the strict technical requirements for such a method to deliver reliable results. Future work on this project will involve testing this method by injecting artificial flares into simulated LSST-like images and testing the method's ability to precisely recover positional offsets and accurately infer flare temperatures using Rubin data products. 

\subsection{Acknowledgements}

This material is based upon work supported by the National Science Foundation under Award Number AST-2308016. 

JRAD acknowledges support from the DiRAC Institute in the Department of Astronomy at the University of Washington. The DiRAC Institute is supported through generous gifts from the Charles and Lisa Simonyi Fund for Arts and Sciences, and the Washington Research Foundation.

The authors acknowledge the support of the Vera C. Rubin Legacy Survey of Space and Time Science Collaborations\footnote{\url{https://www.lsstcorporation.org/science-collaborations}}\par and particularly of the Transient and Variable Star Science Collaboration\footnote{\url{https://lsst-tvssc.github.io/}} (TVS SC) that provided opportunities for collaboration and exchange of ideas and knowledge.

Based on observations obtained with the Samuel Oschin Telescope 48-inch and the 60-inch Telescope at the Palomar Observatory as part of the Zwicky Transient Facility project. ZTF is supported by the National Science Foundation under Grants No. AST-1440341 and AST-2034437 and a collaboration including current partners Caltech, IPAC, the Weizmann Institute for Science, the Oskar Klein Center at Stockholm University, the University of Maryland, Deutsches Elektronen-Synchrotron and Humboldt University, the TANGO Consortium of Taiwan, the University of Wisconsin at Milwaukee, Trinity College Dublin, Lawrence Livermore National Laboratories, IN2P3, University of Warwick, Ruhr University Bochum, Northwestern University and former partners the University of Washington, Los Alamos National Laboratories, and Lawrence Berkeley National Laboratories. Operations are conducted by COO, IPAC, and UW.

This project used data obtained with the Dark Energy Camera
(DECam), which was constructed by the Dark Energy Survey (DES)
collaboration. Funding for the DES Projects has been provided by the
U.S. Department of Energy, the U.S. National Science Foundation,
the Ministry of Science and Education of Spain, the Science and
Technology Facilities Council of the United Kingdom, the Higher
Education Funding Council for England, the National Center for
Supercomputing Applications at the University of Illinois at Urbana Champaign, the Kavli Institute for Cosmological Physics, University
of Chicago, the Center for Cosmology and Astroparticle Physics,
Ohio State University, the Mitchell Institute for Fundamental Physics and Astronomy at Texas A\&M University, Financiadora de Estudos e Projetos, Fundacao Carlos Chagas Filho de Amparo a Pesquisa do Estado do Rio de Janeiro, Conselho Nacional de Desenvolvimento Cientıfico e Tecnologico and the Ministerio da Ciencia, Tecnologia e Inovacao, the Deutsche Forschungsgemeinschaft, and the Collaborating Institutions in the Dark Energy Survey. The Collaborating Institutions are Argonne National Laboratory, the University of California, Santa Cruz, the University of Cambridge, Centro de Investigaciones Energeticas, Medioambientales y Tecnologicas-Madrid, the University of Chicago, University College.

This research has made use of data and/or services provided by the International Astronomical Union's Minor Planet Center.

This research has made use of NASA’s Astrophysics Data System Bibliographic Services.
\\ 
\\
\emph{Software:}  \texttt{astropy} \citep{astropy:2013, astropy:2022},  \texttt{healpy}, \citep{healpix2005},
\texttt{Jupyter} \citep{jupyter}, 
\texttt{MatplotLib} \citep{matplotlib}, \texttt{NumPy} \citep{numpy}, \texttt{pandas} \citep{pandas,pandas2}, 
\texttt{Python3} \citep{python3}.
\texttt{SciPy} \citep{scipy}, and \texttt{rubin\_sim} \citep{peter_yoachim_2023_10028614}.

\bibliography{references, additionalRefs, lsst}{}
\bibliographystyle{aasjournal}



\end{document}